\documentclass{article}

\usepackage{arxiv}

\usepackage{amssymb}
\usepackage{amsmath}
\usepackage{algorithm}
\usepackage{algpseudocode}

\usepackage[hyphens]{url}
\usepackage{float}
\usepackage{mathtools}

\usepackage{booktabs}
\usepackage{bm}

\title{Deep Material Network: Overview, Applications, and Current Directions}

\author{
 Ting-Ju Wei \\
  Department of Civil Engineering\\
  National Taiwan University\\
  Taipei, Taiwan \\
   \And
Wen-Ning Wan \\
  Department of Civil Engineering\\
  National Taiwan University\\
  Taipei, Taiwan \\
   \And
 Chuin-Shan Chen\thanks{Corresponding author. Email: \texttt{dchen@ntu.edu.tw}} \\
  Department of Civil Engineering\\
  Department of Materials Science and Engineering\\
  National Taiwan University\\
  Taipei, Taiwan \\
}

\begin{document}
\maketitle

\begingroup
\renewcommand\thefootnote{}
\footnotetext{This version of the article has been accepted for publication, after peer review (when applicable) and is subject to Springer Nature’s AM terms of use, but is not the Version of Record and does not reflect post-acceptance improvements, or any corrections. The Version of Record is available online at: \url{https://doi.org/10.1007/s42493-026-00146-4}}
\endgroup

\begin{abstract}
The Deep Material Network (DMN) has emerged as a powerful framework for multiscale materials modeling, enabling
efficient and accurate prediction of material behavior across different length scales. Unlike conventional data-driven
approaches, the trainable parameters in DMN possess clear physical interpretations-they encode the geometric characteristics
of representative volume elements (RVEs) rather than serving as purely statistical fitting parameters . By employing
a hierarchical tree structure, DMN learns the homogenization behavior associated with microstructural geometry.
Consequently, it can be trained exclusively on linear elastic datasets while effectively extrapolating to nonlinear responses
during online prediction, making it a highly efficient and scalable approach for multiscale simulations. From a broader
perspective, DMN can be viewed as a physics-informed reduced-order model that captures the essential micromechanical
features governing macroscopic behavior. Its hierarchical formulation provides a compact yet interpretable representation
of the RVE response, significantly reducing computational costs compared to direct numerical simulations. This review
elaborates on the theoretical foundation, training methodology, and recent extensions of DMN, emphasizing its role as a
unifying framework that connects data-driven learning with physically interpretable multiscale modeling.
\end{abstract}

% keywords can be removed

\section{Introduction}\label{sec1}

Multiscale simulation methods are indispensable in computational
mechanics for bridging the gap between microstructural
features and macroscopic material behavior. Many
engineering materials, such as polycrystals, composites,
and porous media, exhibit pronounced heterogeneities that strongly influence their overall mechanical response. Accurately
capturing these effects requires numerical approaches
that incorporate microscale details while maintaining manageable
computational cost.

A widely adopted strategy is the representative volume
element (RVE), which statistically characterizes material
heterogeneity~\cite{hill1965self,feyel2000fe2,geers2010multi,mohammadpour2022multi,lee2021comparison}. By imposing macroscopic boundary
conditions on the RVE, full-field methods like the finite
element method (FEM) or fast Fourier transform (FFT)-
based solvers can resolve detailed internal stress and strain
fields~\cite{temizer2011adaptive,eisenlohr2013spectral,shanthraj2015numerically,vidyasagar2018deformation,lebensohn2020spectral}. FEM provides flexibility for handling complex
geometries, whereas FFT-based solvers exploit spectral
formulations to achieve high computational efficiency
in periodic domains.

Despite their benchmark-level accuracy, full-field simulations
incur a computational cost that scales steeply with
microstructural complexity and nonlinear material behavior.
Such expense retricts their applicability in large-scale
or real-time analyses, motivating the development of physics-
informed surrogate models that can reproduce RVE
responses at substantially reduced computational expense.

Machine-learning-based surrogate models have therefore
emerged as a promising avenue for accelerating multiscale
materials modeling~\cite{zhang2020using,mozaffar2019deep,bishara2023state,frankel2019predicting,chen2021deep,kim2024review,lee2023advancements,chang2023predict, su2023model, yu2022hierarchical}. However, purely data-driven
approaches typically require large training datasets and
often suffer from limited extrapolation capability, necessitating
retraining when microstructural morphology or constitutive
behavior changes.

To address these limitations, the Deep Material Network
(DMN) was introduced as a physics-informed surrogate
model for multiscale simulation~\cite{liu2019deep,liu2019exploring}. In DMN, the
effective response of an RVE is decomposed into multiple
subdomains, each associated with a constituent (base) material.
These subdomains are hierarchically organized within a
binary-tree–based architecture, where each node represents
a local homogenization operation between two child subdomains,
governed by analytical formulations.

Within each tree node, the homogenization problem is
fully characterized by a small set of microstructure-related
parameters, such as volume fractions and stress equilibrium
directions. As a result, the network parameters encode the
underlying microstructural geometry rather than constitutive
nonlinearity, enabling the DMN to be trained solely
using linear elastic stiffness data.

The offline training procedure and numerical implementation
of the DMN were systematically organized and
rigorously analyzed by Srinivas et al.~\cite{srinivas2026rapid}, providing a comprehensive
understanding of the model’s learning mechanism
and computational characteristics. Once trained, DMN
demonstrates remarkable generalization beyond the linearelastic
regime, accurately predicting the first-order nonlinear
material responses with high fidelity~\cite{gajek2020micromechanics}.

This article provides a comprehensive overview of the DMN framework, including its fundamental principles, recent methodological developments, and practical applications. Section~2 introduces the original DMN architecture and its theoretical foundation. Section~3 explores recent extensions, such as multiphysics coupling and enhanced generalization across diverse microstructures. Section~4 highlights key applications, including component-scale multiscale analysis and inverse parameter identification. Finally, Section~5 concludes with a summary
of current insights and an outlook on future directions for
expanding DMN’s scope in multiscale materials modeling.

\section{Deep Material Network}\label{sec2}

This section provides a comprehensive overview of the
original three-dimensional DMN architecture, emphasizing
its hierarchical representation, theoretical foundation,
and computational workflow. It begins with the formulation
of the fundamental building block, followed by the offline
training procedure for learning the homogenization behavior
of RVEs, and the online prediction algorithm that enables
extrapolation to nonlinear material responses. Finally, the
rationale behind the DMN formulation is presented to clarify
its role as a physics-informed reduced-order representation
of microstructure–property relationships.

\subsection{Hierarchical structure of DMN}

\begin{figure}[H]
    \centering
    \includegraphics[width=\textwidth]{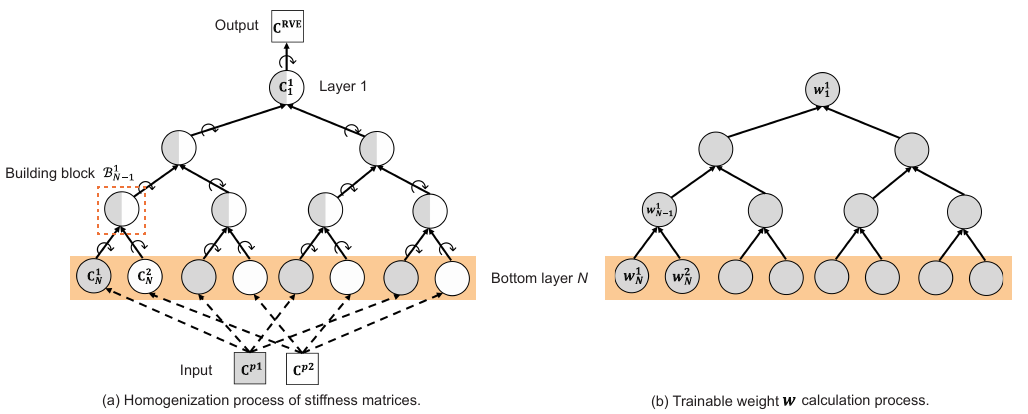}
    \caption{
        Schematic representation of the data flow in the DMN framework.
        (a) Homogenization process of stiffness matrices, where the hierarchical network structure recursively aggregates local stiffness components to compute the effective stiffness $\mathbf{C}^{\text{RVE}}$.
        (b) Calculation of trainable weights $\mathbf{w}$, which parametrize the contributions of different building blocks within the hierarchical architecture. 
    }
    \label{fig:DMN_architecture}
\end{figure}

DMN employs a hierarchical architecture to approximate the response of an RVE. As shown in Fig.~\ref{fig:DMN_architecture}, DMN is structured as a binary tree with $N$ layers. The root node at Layer 1 represents the effective response of the RVE, while the bottom layer ($N$) contains $2^{N-1}$ nodes, each corresponding to a constituent material phase. Additionally, DMN consists of $2^N-1$ fundamental building blocks, each denoted as $\mathcal{B}^k_i$, where $i$ represents the depth of the network and $k$ indexes the building block. 

To quantify the contribution of each constituent material phase, DMN introduces a weighting factor $w^k_i$ at index $k$ and layer $i$ within each building block, as shown in Fig.~\ref{fig:DMN_architecture}(b). The trainable parameters $z^k$ are defined at the bottom nodes, where the weighting factor $w^k$ is activated using the ReLU function~\cite{glorot2011deep}:

\begin{equation}\label{eq:weighting_factor}
    w^k = \text{ReLU}(z^k), \quad k=1,\dots,2^N.
\end{equation}

The weighting factor $w^k$ represents the relative contribution of each bottom node. Due to the hierarchical nature of the DMN, the weighting factors are progressively accumulated across layers and summed at the parent nodes, ensuring a consistent aggregation of contributions:

\begin{equation}\label{eq:weighting_sum_up}
    w^k_i = w^{2k}_{i+1} + w^{2k-1}_{i+1}.
\end{equation}

This hierarchical accumulation enables DMN to capture the interaction between material phases while preserving physical consistency.

\subsection{Mechanistic building block}
This subsection outlines the theoretical foundation of the DMN building block, which serves as the fundamental computational unit for hierarchical homogenization.
The homogenization function within each building block is derived from the principles of linear elasticity; therefore, the offline training data employed in DMN are generated from linear elastic simulations.

Both the stress $\sigma$ and strain $\varepsilon$ tensors are expressed in terms of the Cauchy stress and infinitesimal strain, respectively, using Mandel notation to maintain consistency throughout the formulation:

\begin{equation}
    \sigma = \left[ \sigma_{11},\sigma_{22},\sigma_{33},\sqrt{2}\sigma_{23},\sqrt{2}\sigma_{13},\sqrt{2}\sigma_{12}  \right]^\top \equiv  \left[ \sigma_{1},\sigma_{2},\sigma_{3},\sigma_{4},\sigma_{5},\sigma_{6}  \right]^\top.
\end{equation}
and 
\begin{equation}
    \varepsilon = \left[ \varepsilon_{11},\varepsilon_{22},\varepsilon_{33},\sqrt{2}\varepsilon_{23},\sqrt{2}\varepsilon_{13},\sqrt{2}\varepsilon_{12}  \right]^\top \equiv  \left[ \varepsilon_{1},\varepsilon_{2},\varepsilon_{3},\varepsilon_{4},\varepsilon_{5},\varepsilon_{6}  \right]^\top.
\end{equation}

The macroscopic constitutive relationship is expressed as:
\begin{equation}
    \bar{\sigma} = \bar{\textbf{C}} :\bar{\varepsilon}.
\end{equation}

For a building block comprising two constituent phases, denoted as phase 1 and phase 2, the local stress–strain relationships are given by:

\begin{equation}
    \sigma^{\alpha} = \bar{\textbf{C}}^{\alpha} : \varepsilon^{\alpha}, \quad \alpha = 1,2.
\end{equation}
where $\bar{\textbf{C}}^1$ and $\bar{\textbf{C}}^2$ represent the homogenized stiffness matrices of phases 1 and 2, respectively.

For a building block located at layer $i$ and index $k$, the contribution of each phase is weighted by its corresponding volume fraction $f^{\alpha}$, computed as:

\begin{equation}\label{eq:vf_cal}
    f^{\alpha} = \frac{w^{2k+\alpha-2}_{i+1}}{w^{2k}_{i+1} + w^{2k-1}_{i+1}}, \quad \alpha = 1,2,
\end{equation}
where $w_{i+1}$ denotes the weighting factors associated with the child nodes at the next layer.

These volume fractions quantify the relative contribution of the two phases within the building block.

\begin{figure}[H]
    \centering
    \includegraphics[width=0.5\textwidth]{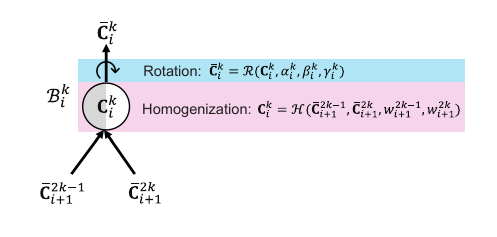}
    \caption{
        Schematic illustration of the homogenization process within a building block $\mathcal{B}^k_i$. 
    }
    \label{fig:homogenization}
\end{figure}

Each building block $\mathcal{B}^k_i$ performs two sequential operations, as illustrated in Fig.~\ref{fig:homogenization}:
\begin{enumerate}
    \item \textit{Stiffness homogenization}: Computes an intermediate homogenized stiffness matrix $\mathbf{C}^k_i$ by aggregating the stiffness tensors of its two child nodes according to micromechanical equilibrium.

    \item \textit{Rotation}: Transforms the intermediate homogenized stiffness $\mathbf{C}^k_i$ into $\bar{\mathbf{C}}^k_i$ through a local coordinate rotation parameterized by the trainable angles $\alpha^k_i$, $\beta^k_i$, and $\gamma^k_i$. This rotation enhances the expressiveness of the homogenization function and enables the network to capture anisotropic behavior.  

\end{enumerate}

\subsubsection{Stiffness homogenization function}

The stiffness homogenization function $\mathcal{H}$ in the DMN is derived from the interfacial equilibrium condition, leading to an analytical expression for the effective stiffness of a two-phase building block.
The interface between the two constituent phases is assumed to be oriented normal to the 3-direction. Under this assumption, strain compatibility is enforced along the 1–2 directions, while traction equilibrium is satisfied along the 3-direction.
A detailed derivation of this formulation can be found in~\cite{liu2019exploring}.
Following this analytical framework, the intermediate homogenized stiffness matrix $\textbf{C}$ can be expressed as:

\begin{equation}\label{eq:homog_function}
    \begin{aligned}
        \textbf{C} &= \mathcal{H}(\bar{\textbf{C}}^1, \bar{\textbf{C}}^2, w^1, w^2) \\
                   &= f_1\bar{\textbf{C}}^1 \bm{\varepsilon}^1 + f_2\bar{\textbf{C}}^2 \bm{\varepsilon}^2 \\
                   &= \bar{\textbf{C}}^2 - f^1\Delta\textbf{C}\textbf{s}^1.
    \end{aligned}
\end{equation}

\noindent where $\Delta\textbf{C} = \bar{\textbf{C}}^1 - \bar{\textbf{C}}^2$ is the stiffness difference between the two phases, and $\textbf{s}^1$ is the strain concentration tensor of phase 1, which determines the strain of phase 1 within the building block.

The strain concentration tensor $\textbf{s}^1$ satisfies:
\begin{equation} \label{eq:strain_localization}
    \bm{\varepsilon}^1=\textbf{s}^1\bm{\varepsilon}   ,\quad \textbf{s}^1_{11} = \textbf{s}^1_{22} = \textbf{s}^1_{66} = 1, \quad \textbf{s}^1_{([3,4,5],:)}=\textbf{s}_{3 \times 6}.
\end{equation}
where the subscripts denote the row and column indices of the respective submatrix.

The submatrix $\textbf{s}_{3\times 6}$ is given by:
\begin{equation}
    \textbf{s}_{3\times 6}= (\hat{\textbf{C}}_{345})^{-1} \begin{bmatrix}
f^2\Delta C_{11} & f^2\Delta C_{23} & \bar{C}^2_{33} & \bar{C}^2_{34} & \bar{C}^2_{35} & f^2\Delta C_{36} \\
f^2\Delta C_{14} & f^2\Delta C_{24} & \bar{C}^2_{34} & \bar{C}^2_{44} & \bar{C}^2_{45} & f^2\Delta C_{46} \\
f^2\Delta C_{15} & f^2\Delta C_{25} & \bar{C}^2_{35} & \bar{C}^2_{45} & \bar{C}^2_{55} & f^2\Delta C_{56}
\end{bmatrix}.
\end{equation}
where
\begin{equation}
    \hat{\textbf{C}} = f^2\bar{\textbf{C}}^1  + f^1 \bar{\textbf{C}}^2
\end{equation}
and
\begin{equation}
    \hat{\textbf{C}}_{345} = \begin{bmatrix}
\hat{{C}}_{33} & \hat{{C}}_{34} & \hat{{C}}_{35}\\
\hat{{C}}_{34} & \hat{{C}}_{44} & \hat{{C}}_{45}\\
\hat{{C}}_{35} & \hat{{C}}_{45} & \hat{{C}}_{55}
\end{bmatrix}.
\end{equation}
Here, the subscripts in $\Delta C$ and $\hat{{C}}$ denote the corresponding row and column indices of the matrices $\Delta\textbf{C}$ and $\hat{\textbf{C}}$, respectively.

\subsubsection{Rotation function}

The rotation function $\mathcal{R}$ in DMN is parameterized using the Tait-Bryan angles $(\alpha, \beta, \gamma)$. The corresponding rotation matrix $\textbf{R}$ is expressed as the product of three elemental rotation matrices:
\begin{equation}\label{eq:rotation_definition}
    \textbf{R}(\alpha, \beta, \gamma) = \textbf{X}(\alpha)\textbf{Y}(\beta)\textbf{Z}(\gamma).
\end{equation}

This rotation matrix is used to transform the intermediate homogenized stiffness matrix $\textbf{C}$, obtained from Eq~\eqref{eq:homog_function}, into its rotated counterpart $\bar{\textbf{C}}$:

\begin{equation}
    \bar{\textbf{C}}=\textbf{R}^{-1}(\alpha, \beta, \gamma)\textbf{C}\textbf{R}(\alpha, \beta, \gamma)
\label{eq:rotation function DMN}
\end{equation}

The individual elemental rotation matrices $\textbf{X}(\alpha)$, $\textbf{Y}(\beta)$, and $\textbf{Z}(\gamma)$ define rotations about the $x$-, $y$-, and $z$-axes, respectively. The subscripts in the following equations indicate the positions of the corresponding matrix components in the elemental rotation matrices, structured in Mandel notation:

\begin{equation}
\begin{split}
    &\textbf{X}_{(1,1)}=1, \textbf{X}_{([2,3,4],[2,3,4])}(\alpha)=\textbf{r}^{p}(\alpha), \textbf{X}_{([5,6],[5,6])}(\alpha)=\textbf{r}^{\nu}(\alpha);\\
    &\textbf{Y}_{(2,2)}=1, \textbf{Y}_{([1,3,5],[1,3,5])}(\beta)=\textbf{r}^{p}(-\beta), \textbf{Y}_{([4,6],[4,6])}(\beta)=\textbf{r}^{\nu}(-\beta);\\
   &\textbf{Z}_{(3,3)}=1, \textbf{Z}_{([1,2,6],[1,2,6])}(\gamma)=\textbf{r}^{p}(\gamma), \textbf{Z}_{([4,5],[4,5])}(\gamma)=\textbf{r}^{\nu}(\gamma);
\end{split}
\end{equation}
Here, $\textbf{r}^{p}$ and $\textbf{r}^{\nu}$ denote the in-plane and out-of-plane rotation matrices, respectively. Given an arbitrary input angle $\theta$, these matrices are defined as:

\begin{equation}
\begin{split}
    &\textbf{r}^{p}(\theta)=\begin{bmatrix}
        \cos^2\theta & \sin^2\theta & \sqrt{2}\sin\theta \cos\theta\\
        \sin^2\theta & \cos^2\theta & -\sqrt{2}\sin\theta \cos\theta\\
        -\sqrt{2}\sin\theta \cos\theta & \sqrt{2}\sin\theta \cos\theta & \cos^2\theta- \sin^2\theta\\
    \end{bmatrix}\\
    &\textbf{r}^{\nu}(\theta)=\begin{bmatrix}
        \cos\theta & -\sin\theta\\
        \sin\theta & \cos\theta\\
    \end{bmatrix}\\
\end{split}
\end{equation}

\subsection{Offline training}

The offline training process in the DMN corresponds to the homogenization procedure, during which the network learns to map the stiffness properties of constituent phases to the homogenized response of an RVE. Using the stiffness matrices of two constituent phases, $\mathbf{C}^{p1}$ and $\mathbf{C}^{p2}$, as input, it computes the homogenized stiffness matrix $\mathbf{C}^{\mathrm{DMN}}$, in accordance with Eq.~\eqref{eq:DMN_homog_overall}. For brevity, the set of trainable parameters in the DMN is collectively denoted as $(\hat{z}, \hat{\alpha}, \hat{\beta}, \hat{\gamma})$.

\begin{equation}\label{eq:DMN_homog_overall}
    \mathbf{C}^{\text{DMN}} = \mathcal{DMN}(\mathbf{C}^{p1}, \mathbf{C}^{p2}, z^{k=1,...,2^{N-1}},\alpha ^{k=1,...,2^{i-1}}_{i=1,...,N}, \beta ^{k=1,...,2^{i-1}}_{i=1,...,N}, \gamma ^{k=1,...,2^{i-1}}_{i=1,...,N})
\end{equation}

The training dataset consists of stiffness matrices, where each sample is represented as a triplet $(\mathbf{C}^{p1}, \mathbf{C}^{p2}, \mathbf{C}^{\text{DNS}})$.
Here, $\mathbf{C}^{p1}$ and $\mathbf{C}^{p2}$ denote the stiffness matrices of the individual phases, while $\mathbf{C}^{\text{DNS}}$ represents the homogenized stiffness matrix obtained from direct numerical simulations (DNS).
The goal is to exploit the contrast between $\mathbf{C}^{p1}$ and $\mathbf{C}^{p2}$, allowing the DMN to infer the underlying geometric characteristics of the RVE through data-driven learning.

The loss function $\mathcal{L}$ is defined as
\begin{equation}
    \mathcal{L} = \frac{1}{2N_s}\sum_{s} \mathcal{L}_{\text{stiff}} + \lambda(\sum_{j}(\text{ReLU}(z^j)-2^{N-2}))^2,
\end{equation}
where $\lambda$ is a regularization coefficient and $s$ denotes the sample index.

The stiffness-based loss term $\mathcal{L}_{\text{stiff}}$ is expressed as
\begin{equation}
    \mathcal{L}_{\text{stiff}} = \frac{\left\| \mathbf{C}_s^{\text{DNS}} -\mathcal{DMN}(\mathbf{C}_s^{p1}, \mathbf{C}_s^{p2}, \hat{z}, \hat{\alpha}, \hat{\beta}, \hat{\gamma}) \right\|^2}{\left\| \mathbf{C}_s^{\text{DNS}} \right\|^2} 
\end{equation}
During training, the dataset is divided into mini-batches, and the parameters $(\hat{z}, \hat{\alpha}, \hat{\beta}, \hat{\gamma})$ are updated using stochastic gradient descent (SGD)~\cite{bottou2010large}.

The forward propagation of the DMN homogenization process proceeds as follows.
Initially, the stiffness matrices are assigned to the material nodes at the bottom layer according to

\begin{equation}\label{eq:stiffness_assigned}
    {\mathbf{C}}^{k}_N = 
    \begin{cases}
        \mathbf{C}^{p1}, & k=1,3,...,2^N-1 \\
        \mathbf{C}^{p2}, & k=2,4,...,2^N
    \end{cases}.
\end{equation}
where $N$ denotes the total number of layers in the binary-tree network.

From the bottom layer upward, the homogenization proceeds hierarchically through the building blocks.
At each block, Eq.~\eqref{eq:homog_function} is applied for \textit{stiffness homogenization}, while Eq.~\eqref{eq:rotation function DMN} governs the \textit{rotation transformation}.
This recursive process ensures that stress–strain equilibrium is satisfied at all intermediate scales.
Finally, the homogenized stiffness matrix at the root node, ${\mathbf{C}}^1_1$, represents the macroscopic stiffness of the RVE, expressed as

\begin{equation}
    \mathbf{C}^{\text{DMN}} = {\mathbf{C}}^1_1.
\end{equation}

\subsection{Online prediction}

The online prediction phase extends the trained DMN to nonlinear material regimes by introducing a residual strain field at the bottom-layer material units $\mathcal{B}^j_N$ ($j = 1, \dots, 2^{N-1}$).
Each unit acts as an independent material point, characterized by its strain $\bm{\varepsilon}_N^j$, stress $\bm{\sigma}_N^j$, internal variables $\bm{\beta}_N^j$, and residual strain $\delta\bm{\varepsilon}_N^j$.

At the root node $\mathcal{B}^1_1$, the macroscopic stress–strain relationship is expressed as

\begin{equation}\label{eq:root_ss}
    \Delta\bm{\varepsilon}^{\text{DMN}}  = {\textbf{C}^{\text{DMN}}}^{-1} \Delta \bm{\sigma}^{\text{DMN}} + \delta\bm{\varepsilon}^{\text{DMN}},
\end{equation}
where $\mathbf{C}^{\text{DMN}}$ represents the homogenized tangent stiffness of the network, and $\delta\bm{\varepsilon}^{\text{DMN}}$ denotes the homogenized accumulated residual strain.

\begin{figure}[H]
    \centering
    \includegraphics[width=0.7\textwidth]{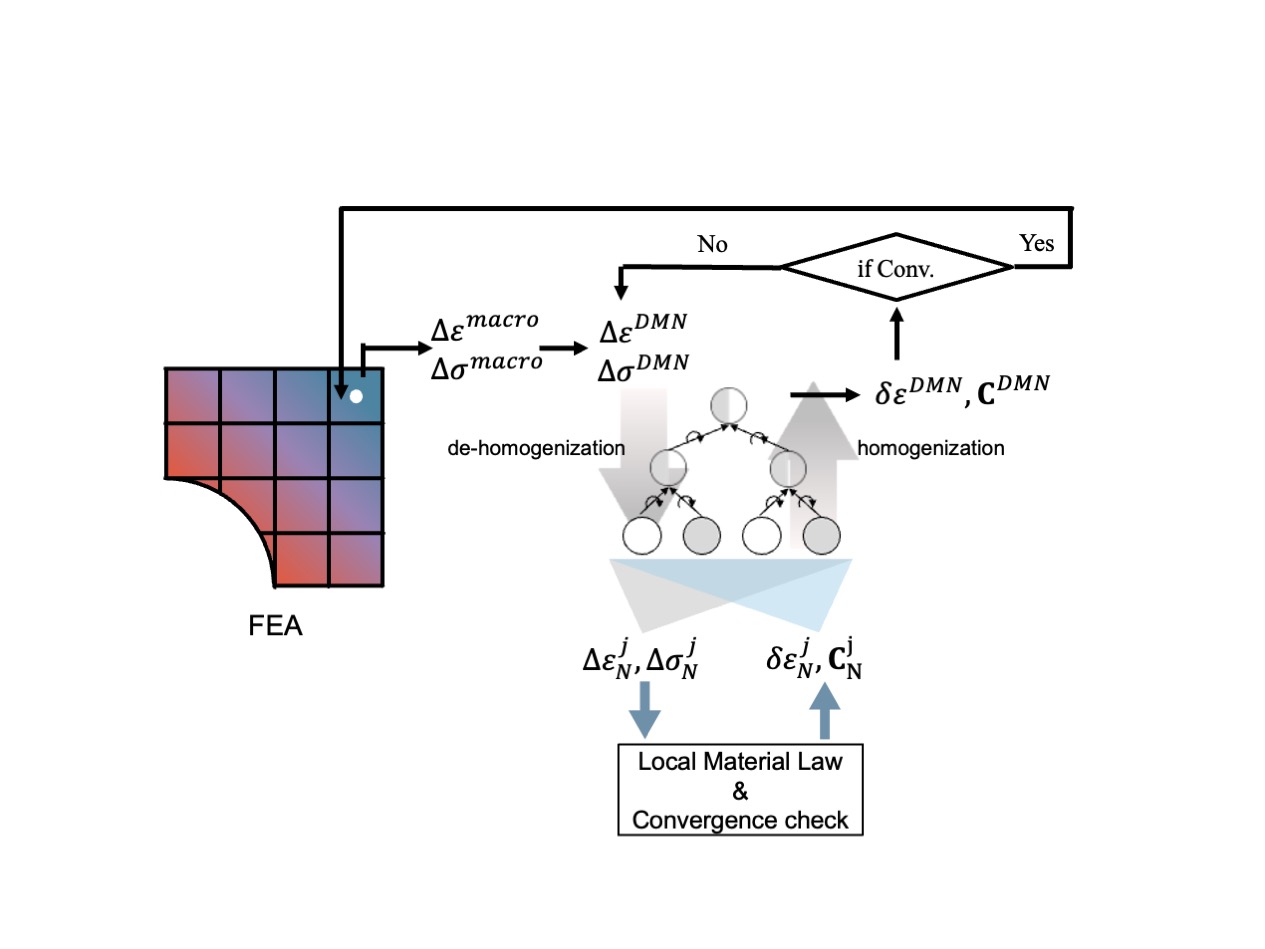}
    \caption{ Schematic representation of DMN in the online prediction phase.
    }
    \label{fig:DMN online}
\end{figure}

Given an imposed macroscopic loading condition, the objective of the DMN is to achieve global equilibrium between the local material responses and the applied load.
This is accomplished through a Newton–Raphson iterative procedure alternating between homogenization and de-homogenization processes, as illustrated in Fig.~\ref{fig:DMN online}.

During the homogenization step, the residual strain $\delta\bm{\varepsilon}^{\text{DMN}}$ and the homogenized stiffness matrix $\textbf{C}^{\text{DMN}}$ are obtained by upscaling information from the bottom nodes to the root node.
Based on the macroscopic loading condition, the overall incremental strain $\Delta\bm{\varepsilon}^{\text{DMN}}$ and stress $\Delta\bm{\sigma}^{\text{DMN}}$ are then determined.

In the de-homogenization step, $\Delta\bm{\varepsilon}^{\text{DMN}}$ and $\Delta\bm{\sigma}^{\text{DMN}}$ propagate back to the bottom nodes, where the local material law is evaluated. At each iteration $i$, the bottom node strains update as $\Delta\bm{\varepsilon}_N^{j,\text{iter}=i}$. Convergence is reached when the relative difference between $\Delta\bm{\varepsilon}_N^{j,\text{iter}=i}$ and $\Delta\bm{\varepsilon}_N^{j,\text{iter}=i-1}$ falls below a predefined threshold; otherwise, the updated strain continues in the next iteration.

At the bottom layer $N$, each node $\mathcal{B}^j_N$ follows the local constitutive relation:
\begin{equation}
    \Delta\bm{\sigma}^j_N = \Delta\bm{\sigma}^j_N(\Delta\bm{\varepsilon}^j_N, \bm{\varepsilon}^j_N,\bm{\sigma}^j_N, \bm{\beta}^j_N)
\end{equation}
and 
\begin{equation}
    \textbf{C}^j_N = \textbf{C}^j_N(\Delta\bm{\varepsilon}^j_N, \bm{\varepsilon}^j_N,\bm{\sigma}^j_N, \bm{\beta}^j_N)
\end{equation}

From these relations, both $\Delta\bm{\sigma}^j_N$ and $\textbf{C}^j_N$ are determined, allowing the residual strain at each bottom node to be expressed as

\begin{equation}
    \delta\bm{\varepsilon}^j_N = \Delta\bm{\varepsilon}^j_N - {\textbf{D}^j_N} \Delta \bm{\sigma}^j_N
\end{equation}
where $\textbf{D}^j_N = (\textbf{C}^j_N)^{-1}$ denotes the local compliance matrix.

The computed $\delta\bm{\varepsilon}^j_N$ and $\textbf{C}^j_N$ are then homogenized to the root node $\mathcal{B}^1_1$ using the homogenization function defined in Eq.~\eqref{eq:homog_function} and the rotation operation in Eq.~\eqref{eq:rotation function DMN}.

Similarly, the residual strain undergoes an intermediate homogenization step, followed by a rotation operation, with updates applied layer by layer. The intermediate homogenized residual strain is computed as:

\begin{equation}\label{eq:residual_h1}
\begin{split}
&\delta{\bm{\varepsilon}} = f_1\delta\bm{\varepsilon}^{1}+f_2\delta\bm{\varepsilon}^{2}+\\
&f_1f_2(\begin{bmatrix}
 D^{1}_{11} &  D^{1}_{12} &  D^{1}_{16}\\
 D^{1}_{21} &  D^{1}_{22} &  D^{1}_{26}\\
 D^{1}_{31} &  D^{1}_{32} &  D^{1}_{36}\\
 D^{1}_{41} &  D^{1}_{42} &  D^{1}_{46}\\
 D^{1}_{51} &  D^{1}_{52} &  D^{1}_{56}\\
 D^{1}_{61} &  D^{1}_{62} &  D^{1}_{66}\\
\end{bmatrix}-\begin{bmatrix}
 D^{2}_{11} &  D^{2}_{12} &  D^{2}_{16}\\
 D^{2}_{21} &  D^{2}_{22} &  D^{2}_{26}\\
 D^{2}_{31} &  D^{2}_{32} &  D^{2}_{36} \\
 D^{2}_{41} &  D^{2}_{42} &  D^{2}_{46} \\
 D^{2}_{51} &  D^{2}_{52} &  D^{2}_{56} \\
 D^{2}_{61} &  D^{2}_{62} &  D^{2}_{66} \\
\end{bmatrix})\\
&\begin{bmatrix}
f_1 D^{2}_{11}+f_2 D^{1}_{11} & f_1 D^{2}_{12}+f_2 D^{1}_{12} & f_1 D^{2}_{16}+f_2 D^{1}_{16}\\
f_1 D^{2}_{21}+f_2 D^{1}_{21} & f_1 D^{2}_{22}+f_2 D^{1}_{22} & f_1 D^{2}_{26}+f_2 D^{1}_{26}\\
f_1 D^{2}_{61}+f_2 D^{1}_{61} & f_1 D^{2}_{62}+f_2 D^{1}_{62} & f_1 D^{2}_{66}+f_2 D^{1}_{66}\\ 
\end{bmatrix}^{-1}\begin{bmatrix}
    \delta\varepsilon^{2}_1 - \delta\varepsilon^{1}_1 \\
    \delta\varepsilon^{2}_2 - \delta\varepsilon^{1}_2 \\
    \delta\varepsilon^{2}_6 - \delta\varepsilon^{1}_6\\
\end{bmatrix}
\end{split}
\end{equation}
where superscripts $^{1}$, $^{2}$ denote phase 1 and phase 2, while subscripts indicate the compliance matrix indices or vector components.

The rotated residual strain is then computed as:

\begin{equation}\label{eq:residual_rot}
    \delta\bar{\bm{\varepsilon}} = \textbf{R}(\alpha, \beta, \gamma)\delta{\bm{\varepsilon}}
\end{equation}

In the 3D DMN study by Liu and Wu, a near-linear relationship between the online computational time and the number of active bottom-layer nodes ($N_a$) was observed for a particle-reinforced composite RVE. In their study, for a DMN depth of $N=8$ (corresponding to $N_a=28$), the DMN online prediction required $6.0$ s on a single CPU core and achieved an approximately $8100\times$ speedup over DNS in terms of CPU time~\cite{liu2019exploring}.

Regarding nonlinear accuracy, it is generally observed that increasing the DMN depth $N$ enhances stress prediction accuracy by improving the representational capacity of the hierarchical network. Nevertheless, for a fixed network depth, the achievable accuracy is strongly influenced by the degree and nature of nonlinearity in the underlying constituent material models.
Reported studies indicate that, for the same DMN depth, the maximum stress prediction error can vary substantially across different nonlinear material behaviors, ranging from below $1\%$ for elasto-plastic responses to approximately $7\%$ for viscoelastic responses. To mitigate this sensitivity, several works have proposed modified offline training strategies, such as tailoring the loss function to emphasize nonlinear response characteristics, which have been shown to reduce viscoelastic prediction errors to below $2\%$~\cite{gajek2023material, dey2022training}.

\subsection{Rationale beyond the DMNs}

The effectiveness of DMNs arises from their physics-informed architecture, which intrinsically enforces thermodynamic consistency and material stability.
Unlike purely data-driven neural networks that infer constitutive behavior solely from data, DMNs embed fundamental micromechanical principles directly within their hierarchical topology.
This architectural bias ensures essential physical conditions—such as monotonic stress–strain behavior, positive energy dissipation, and stable material evolution—without the need for external penalty terms or artificial constraints~\cite{gajek2020micromechanics}.

Several key mechanisms enable DMNs to maintain strong physical consistency and robust performance even beyond their training domain~\cite{gajek2020micromechanics}:
\begin{itemize}
    \item Monotonic stress-strain response: Each DMN building block is derived from interfacial equilibrium conditions. When the constituent materials at the bottom nodes exhibit monotonic stress–strain behavior, this property is preserved throughout the entire network. Consequently, the DMN response remains free from artificial softening or nonphysical instabilities, ensuring both numerical robustness and physically realistic deformation paths.
    
    \item Convexity and stability of the energy: The Helmholtz free energy of each building block is formulated as a weighted combination of the free energies of its child phases. This construction ensures the convexity of the global energy functional, guaranteeing a unique, stable, and physically admissible constitutive response. Such convexity directly contributes to the numerical stability of nonlinear simulations involving DMN-based materials.

    \item Inherited dissipation inequality: The DMN framework inherently satisfies the second law of thermodynamics. Since each constituent phase enforces non-negative energy dissipation, the hierarchical aggregation ensures that the overall network strictly preserves the dissipation inequality. As a result, all inelastic processes within the DMN lead to a net positive (or zero) dissipation, thereby precluding any artificial energy generation.
    
    \item Intrinsic enforcement of physics: Instead of relying on externally imposed constraints or penalty terms (e.g., embedding governing equations into the loss function), DMNs encode thermodynamic laws directly within their structural formulation. This inductive bias ensures that equilibrium, stability, and dissipation are satisfied throughout the network, preserving thermodynamic consistency even when operating far beyond the range of training data.

\end{itemize}

Overall, DMNs provide a robust framework for tackling multiscale and nonlinear material problems by embedding physical principles into every building block.
This intrinsic physics enforcement ensures monotonicity, stable energy evolution, and non-negative dissipation, thereby guaranteeing both thermodynamic soundness and numerical stability.
Consequently, DMNs deliver reliable predictions within the training domain and demonstrate remarkable extrapolation capability beyond it.

\section{Current Directions}\label{sec3}
\subsection{Advancements in DMN methodologies} 

Recent advancements in DMN methodologies have significantly enhanced their computational efficiency, physical interpretability, and applicability across diverse material systems. Table~\ref{tab:dmn_advancements} provides an overview of key developments in DMN-based models, highlighting their primary objectives and relevant references.

\begin{table}[h!]
\centering
\caption{Recent advancements in DMN-based models and their objectives.}
\label{tab:dmn_advancements}
\begin{tabular}{@{}lll@{}} 
\toprule
\textbf{Model} & \textbf{Key Objective} & \textbf{Reference} \\
\midrule
Rotation-free DMN & Extend DMN to multiphase materials and eliminate rotational DOFs\footnotemark[1]  & \cite{gajek2020micromechanics} \\
IMN & Extend DMN to porous materials with enforced Hill-Mandel condition & \cite{noels2022micromechanics, noels2022interaction} \\
ODMN & Extend DMN to polycrystalline materials with texture evolution & \cite{wei2025orientation} \\
MIpDMN & Extend DMN to incorporate thermal (conductivity \& expansion) effects & \cite{li2024micromechanics} \\
Thermomechanical DMN & Extend DMN to incorporate thermal (expansion) effects & \cite{shin2024deep} \\
Thermal DMN & Extend DMN to incorporate thermal (conductivity) effects & \cite{shin2024deep2} \\
DMN with Damage Effect & Extend DMN to integrate cohesive networks for damage modeling & \cite{liu2020deep, liu2021cell} \\
FDMN & Extend DMN to model non-Newtonian fluid dynamics & \cite{sterr2024deep} \\
\bottomrule
\end{tabular}
\footnotetext[1]{DOFs: degrees of freedoms}
\end{table}

\newpage

\subsubsection{Rotation-free DMN} 

Gajek et al. extended the DMN framework by identifying that rotational DOFs at the bottom nodes of undirected composite materials are redundant~\cite{gajek2020micromechanics}. Their analysis, based on Volterra series approximations and multiple-input multiple-output (MIMO) dynamical system frameworks, demonstrated that these DOFs have negligible influence on the overall material response.

By eliminating these unnecessary DOFs, the rotation-free DMN significantly improves computational efficiency while maintaining predictive accuracy. Furthermore, this streamlined formulation naturally extends to multiphase materials, making it a scalable and versatile framework for modeling complex microstructures~\cite{gajek2020micromechanics}.

\subsubsection{Interaction-based material network} 

Building upon the rotation-free DMN, Van Dung Nguyen and Ludovic Noels introduced the interaction-based material network (IMN) to extend DMN's applicability to porous materials~\cite{noels2022micromechanics, noels2022interaction}.

\begin{figure}[H]
    \centering
    \includegraphics[width=0.5\textwidth]{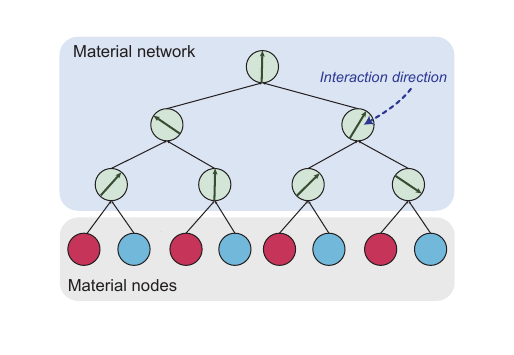}
    \caption{
    Schematic illustration of the IMN framework, which consists of a material network and a set of material nodes.
}

    \label{fig:IMN}
\end{figure}

IMN reformulates DMN by explicitly distinguishing material nodes from the material network, as shown in Fig.~\ref{fig:IMN}. In this framework, the bottom nodes of DMN are treated as independent material nodes, while the remaining hierarchical structure constitutes the material network, where each tree node represents an interaction mechanism. An IMN consists of $N$ material nodes and $M$ tree nodes, which can be interpreted as modeling an RVE with $N$ independent material units, grouped into $M$ interaction sets, each of which must satisfy the Hill-Mandel condition. Each interaction mechanism within the material network is characterized by:
\begin{itemize}
    \item An interaction direction $\mathbf{G}^j$, corresponding to the force-equilibrium direction.
    \item An interaction incompatibility $\mathbf{a}^j$, representing deformation fluctuations.
\end{itemize}

A key component of IMN in online prediction is its interaction mapping, which governs the distribution of deformation gradients across material nodes. This mapping is defined as
\begin{equation}\label{eq:interaction_mapping}
   \mathbf{F}^i = \bar{\mathbf{F}} + \sum_{j=0}^{M} \alpha^{i,j} \mathbf{a}^j \otimes \mathbf{G}^j, \quad \text{for } i = 0, 1, \dots, N-1 
\end{equation}
where $i$ indexes the material nodes, and $j$ indexes the interaction mechanisms. From Eq.~\eqref{eq:interaction_mapping}, the fluctuation part of the deformation gradient is decomposed into $M$ interaction modes, each governed by a specific interaction mechanism represented by $\mathbf{a}^j \otimes \mathbf{G}^j$.

Beyond its theoretical foundation, IMN introduces a significant computational advantage over traditional DMN. Unlike DMN, which requires layer-by-layer reconstruction during online prediction, IMN's interaction mapping reformulates de-homogenization as a matrix operation, significantly improving computational efficiency. Studies have demonstrated that IMN achieves substantial speedup in both offline training and online prediction~\cite{wan2024decoding}.

\subsubsection{Orientation-aware interaction-based DMN}

\begin{figure}[H]
    \centering
    \includegraphics[width=0.5\textwidth]{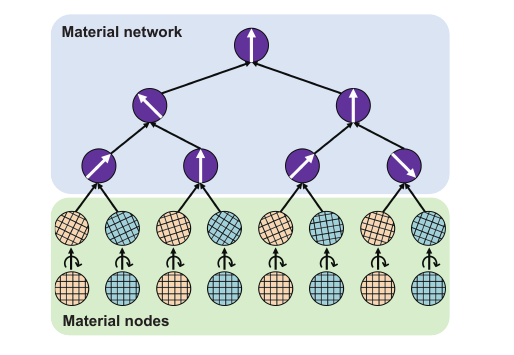}
        \caption{
        Schematic illustration of the ODMN framework. Each material node is associated with trainable rotation angles, encoding local orientation information within the RVE.
    }
    
    \label{fig:ODMN}
\end{figure}

Building upon the IMN framework, recent studies have introduced the orientation-aware interaction-based DMN (ODMN), which incorporates an orientation-aware mechanism at the material nodes, making it applicable to multiphase polycrystalline material systems~\cite{wei2025orientation}, as shown in Fig.~\ref{fig:ODMN}. This mechanism introduces three trainable parameters at each material node: the Tait–Bryan angles \( \alpha \), \( \beta \), and \( \gamma \). These angles define a set of elementary rotation matrices, which collectively govern the rotation of each material node.

During online prediction, the rotation matrix $\mathbf{R}^i$ for a material node index $i$ is constructed using the trainable angles as follows:

\begin{equation}
    \mathbf{R}^i =
    \begin{bmatrix}
    1 & 0 & 0 \\
    0 & \cos \alpha^i & -\sin \alpha^i \\
    0 & \sin \alpha^i & \cos \alpha^i \\
    \end{bmatrix}
    \begin{bmatrix}
    \cos \beta^i & 0 & \sin \beta^i \\
    0 & 1 & 0 \\
    -\sin \beta^i & 0 & \cos \beta^i \\
    \end{bmatrix}
    \begin{bmatrix}
    \cos \gamma^i & -\sin \gamma^i & 0 \\
    \sin \gamma^i & \cos \gamma^i & 0 \\
    0 & 0 & 1 \\
    \end{bmatrix}
\end{equation}

At the initial loading state ($t=0$), the elastic and plastic deformation gradients are initialized as:

\begin{equation}\label{eq: FeFp_init}
    \mathbf{F}^i_e(t=0) = \mathbf{R}^i, \quad
    \mathbf{F}^i_p(t=0) = \mathbf{{R}^i}^{-1}
\end{equation}

For $t>0$, the updated rotation matrix $\mathbf{R}^i_t$ at material node $i$ can be obtained via the polar decomposition of the elastic deformation gradient:

\begin{equation} \label{eq:polar decomposition}
    \mathbf{F}^i_e(t>0) = \mathbf{R}^i_t \mathbf{U}^i_t
\end{equation}

The extracted rotation matrix $\mathbf{R}^i_t$ and $w^i$ are then used to reconstruct the orientation distribution function (ODF), enabling the characterization of crystallographic texture evolution.

While the original DMN can be applied to polycrystalline materials, a key limitation arises from the entanglement of bottom node rotation angles with the rotation angles of hierarchical building blocks within the homogenization function. In the original DMN, each building block is assumed to rotate throughout both homogenization and de-homogenization. Consequently, any change in the bottom node rotation disrupts the force-equilibrium condition of the DMN, leading to simulation instability and non-convergence during online prediction.

In contrast, ODMN decouples material rotation from building block rotations. Rotation does not occur in the building blocks; rather, force-equilibrium directions are established. This distinction eliminates the need for rotation-related operations during both homogenization and de-homogenization in online prediction. As a result, ODMN allows material nodes to undergo deformation-induced rigid-body rotation while preserving the Hill-Mandel condition, thereby enabling the prediction of texture evolution in polycrystalline materials.

\subsubsection{Micromechanics-informed parametric DMN } 

The DMN framework has been further extended to incorporate anisotropic thermal conductivity and thermal expansion in the micromechanics-informed parametric DMN (MIpDMN)~\cite{li2024micromechanics}. In the original DMN building block, the laminate homogenization function for stiffness is expressed as:

\begin{equation}
    \bar{\textbf{C}} = \text{Lam}_{\mathbb{C}}(\textbf{C}^{p1}, \textbf{C}^{p2}).
\end{equation}

Similarly, the effective thermal conductivity $\bar{k}$ follows:

\begin{equation}
    \bar{{k}} = \text{Lam}_{k}({k}^{p1}, {k}^{p2}).
\end{equation}

The laminate thermal homogenization function is governed by Fourier's law:
\begin{equation}
    \vec{q} = -k\nabla T.
\end{equation}

To derive the effective thermal conductivity, the heat flux $\vec{q}$ and the temperature gradient $\nabla T$ are decomposed into their tangential and normal components, denoted by superscripts $^t$ and $^n$, respectively. At the interface, the continuity conditions impose:
\begin{equation}
    \begin{bmatrix}
\nabla T^t_2 \\
\vec{q}^n_1
\end{bmatrix}
=
\begin{bmatrix}
\nabla T^t_2  \\
\vec{q}^n_2
\end{bmatrix}
\end{equation}

By solving for the effective thermal conductivity, we obtain:

\begin{equation}
    \bar{k}= f^1(k_1-k_2)\textbf{A} + k_2,\quad \textbf{A}=\tilde{k}^{-1}\tilde{k}_2, \quad \tilde{k}=(1-f^1)\tilde{k}_1 + f^1 \tilde{k}_2
\end{equation}
where 
\begin{equation}
    \tilde{k}_\alpha = \begin{bmatrix}
\mathbb{I}_{2 \times 2} & \textbf{0}_{2 \times 1} \\
k^{nt}_\alpha & k^{nn}_\alpha
\end{bmatrix}, \quad \alpha=1,2
\end{equation}

Beyond thermal conductivity, MIpDMN also incorporates effective stiffness homogenization while accounting for thermal expansion. The corresponding laminate homogenization function is expressed as:

\begin{equation}
    (\bar{\textbf{C}},\bar{\alpha}) = \text{Lam}_{\mathbb{C}, 
\alpha}(\textbf{C}^{p1}, \textbf{C}^{p2}, \alpha_1, \alpha_2).
\end{equation}

Assuming a uniform temperature difference $\Delta T$ and a traction vector $\bar{\sigma} \cdot \vec{n}$ on the RVE boundary, the thermoelastic constitutive equation yields:
\begin{equation}
    \varepsilon_1 = \varepsilon_2 = \mathbb{S}_1 \bar{\sigma} + \alpha_1 \Delta T=\mathbb{S}_2 \bar{\sigma} + \alpha_2 \Delta T
\end{equation}
where $\mathbb{S}_1$ and $\mathbb{S}_2$ represent the compliance matrices of phases 1 and 2, respectively.

By further derivation, the effective thermal expansion coefficient is given by:
\begin{equation}
    \bar{\alpha} = \alpha_1+(\bar{\mathbb{S}} - \mathbb{S}_1)(\mathbb{S}_1 - \mathbb{S}_2)^{-1}(\alpha_1 - \alpha_2).
\end{equation}

Furthermore, MIpDMN enhances the adaptability of DMN to various microstructures, as demonstrated in a later section. To validate the proposed MIpDMN framework, an analysis was conducted of its predictions for the effective thermal conductivity and effective coefficient of thermal expansion in ellipsoidal inclusion composites, considering varying fiber volume fractions and aspect ratios. The results confirm that MIpDMN accurately predicts both effective thermal conductivity and the effective coefficient of thermal expansion across different microstructures.

\subsubsection{Thermomechanical DMN} 

\begin{figure}[H]
    \centering
    \includegraphics[width=0.7\textwidth]{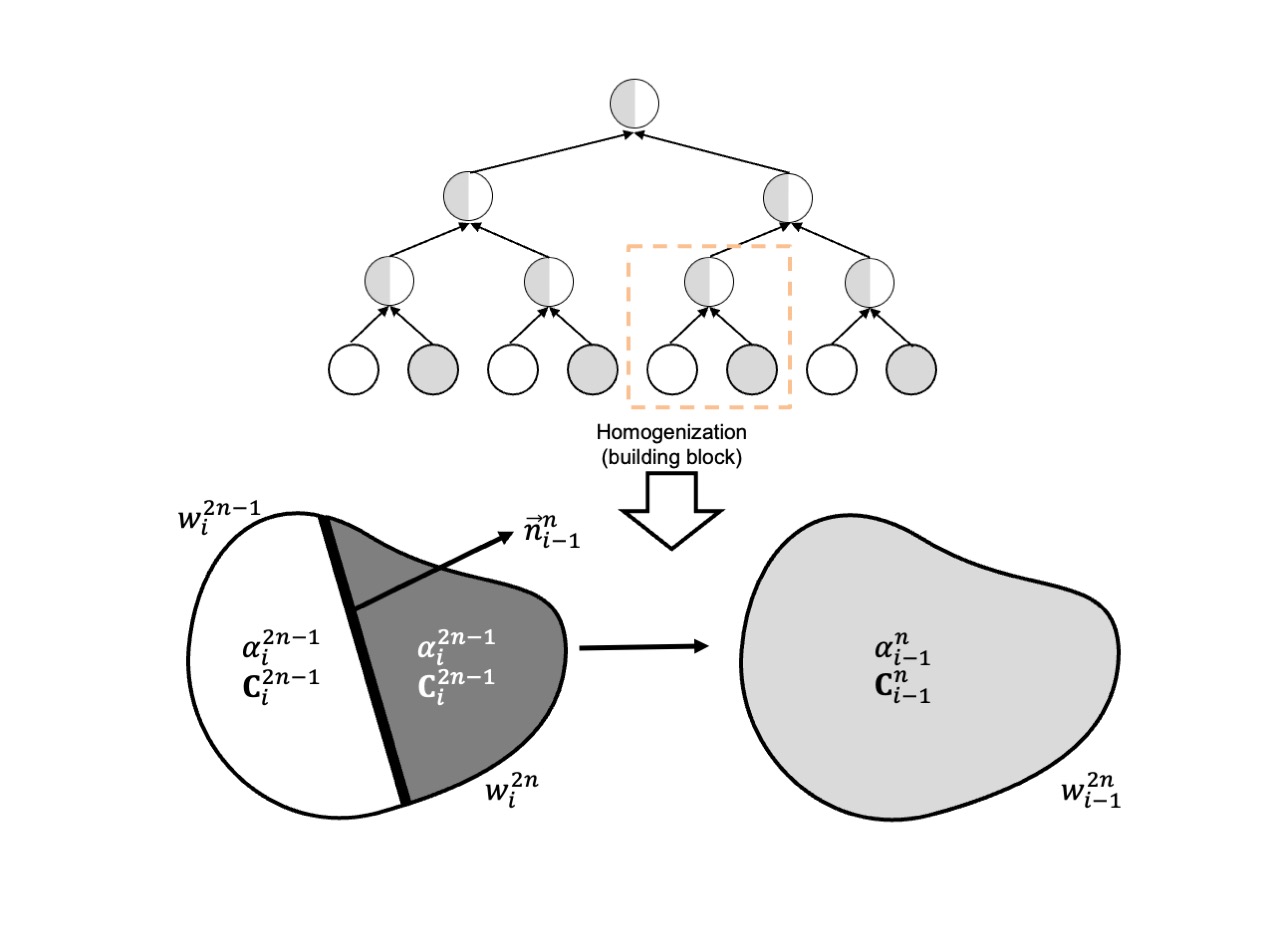}
    \caption{ Schematic illustration of thermomechanical DMN framework.
    }
    \label{fig:thermal expansion}
\end{figure}

Building upon the isothermal DMN framework~\cite{noels2022micromechanics, gajek2020micromechanics, shin2023deep}, Shin et al.~\cite{shin2024deep} extended the methodology to thermomechanical problems by incorporating thermal expansion effects while ensuring energy conservation. This extension is achieved by enforcing interface traction continuity within each building block and modifying the Hill--Mandel condition to account for temperature gradients, as shown in Fig.~\ref{fig:thermal expansion}.

For each building block, the continuity of interface traction is given by
\begin{equation}\label{eq:0}
    \mathbf{H}^\top_h \bigl(\boldsymbol{\sigma}_1 - \boldsymbol{\sigma}_2 \bigr) 
    \;=\; 
    \mathbf{0},
\end{equation}
where \(\mathbf{H}^\top_h\) is a \(3 \times 6\) matrix associated with the interface normal direction. The homogenized strain and stress satisfy the averaging theorem:
\begin{equation}\label{eq:1}
\bar{\boldsymbol{\varepsilon}} \;=\; f_1\,\boldsymbol{\varepsilon}_1 + f_2\,\boldsymbol{\varepsilon}_2,
\quad\quad
\bar{\boldsymbol{\sigma}} \;=\; f_1\,\boldsymbol{\sigma}_1 + f_2\,\boldsymbol{\sigma}_2,
\end{equation}
where \(f_1\) and \(f_2\) are the volume fractions of the two phases. To incorporate thermal effects, the Hill--Mandel condition is extended following Levin's theorem:
\begin{equation}\label{eq:2}
(\bar{\boldsymbol{\varepsilon}} - \bar{\boldsymbol{\alpha}}\,\Delta T)^\top \bar{\boldsymbol{\sigma}}
\;=\;
f_1\,
\bigl(\boldsymbol{\varepsilon}_1 - \boldsymbol{\alpha}_1\,\Delta T \bigr)^\top \boldsymbol{\sigma}_1
\;+\;
f_2\,
\bigl(\boldsymbol{\varepsilon}_2 - \boldsymbol{\alpha}_2\,\Delta T \bigr)^\top \boldsymbol{\sigma}_2.
\end{equation}
Rearranging Eq.~\eqref{eq:2} yields
\begin{equation}\label{eq:3}
    f_1f_2\,(\boldsymbol{\varepsilon}_1 - \boldsymbol{\varepsilon}_2)^\top
    (\boldsymbol{\sigma}_1 - \boldsymbol{\sigma}_2)
    \;=\;
    \Bigl[
        f_1\,(\boldsymbol{\alpha}_1 - \bar{\boldsymbol{\alpha}})^\top \boldsymbol{\sigma}_1
        \;+\;
        f_2\,(\boldsymbol{\alpha}_2 - \bar{\boldsymbol{\alpha}})^\top \boldsymbol{\sigma}_2
    \Bigr] \,\Delta T.
\end{equation}

The homogenization problem can be divided into two subproblems: one under isothermal conditions ($\Delta T$) with a homogeneous traction boundary condition, and the other accounting for traction-free thermal variations.

Under isothermal conditions (\(\Delta T = 0\)), Eq.~\eqref{eq:3} simplifies to
\begin{equation}\label{eq:4}
    f_1f_2\,(\boldsymbol{\varepsilon}_1 - \boldsymbol{\varepsilon}_2)^\top
    (\boldsymbol{\sigma}_1 - \boldsymbol{\sigma}_2)
    =
    0.
\end{equation}
Using Eq.\eqref{eq:0}, this leads to
\begin{align}\label{eq:44}
    (\boldsymbol{\varepsilon}_1 - \boldsymbol{\varepsilon}_2)^\top 
    &= 
    \frac{1}{f_1 f_2}\,(\mathbf{H}\,\mathbf{b})^\top, \notag\\[6pt]
    \boldsymbol{\varepsilon}_1 
    &= 
    \bar{\varepsilon} + \frac{\mathbf{H}\,\mathbf{b}}{f_1},\\[6pt]
    \boldsymbol{\varepsilon}_2 
    &= 
    \bar{\varepsilon} - \frac{\mathbf{H}\,\mathbf{b}}{f_2},\notag
\end{align}
where \(\mathbf{b}\) is an arbitrary vector, and \(\bar{\varepsilon}\) is the homogenized strain. 

In the presence of a temperature variation, Eq.~\eqref{eq:3} simplifies to
\begin{equation}\label{eq:55}
    f_1 \bigl(\boldsymbol{\alpha}_1 - \bar{\boldsymbol{\alpha}}\bigr)^\top \boldsymbol{\sigma}_1
    \;+\;
    f_2 \bigl(\boldsymbol{\alpha}_2 - \bar{\boldsymbol{\alpha}}\bigr)^\top \boldsymbol{\sigma}_2
    \;=\;
    0,
\end{equation}
where \(\Delta T\) can take an arbitrary value. Using Eqs.\eqref{eq:44} and \eqref{eq:55}, the linear homogenization within each building block is given by
\begin{equation}
    \bar{\mathbf{C}}
    \;=\;
    f_1\,\mathbf{C}_1
    \;+\;
    f_2\,\mathbf{C}_2
    \;+\;
    (\mathbf{C}_1 - \mathbf{C}_2)\,\mathbf{H}\,\mathbf{X}_1,
\end{equation}
\begin{equation}
    \bar{\boldsymbol{\alpha}}
    \;=\;
    \bar{\mathbf{C}}^{-1}\,
    \Bigl[
        (\mathbf{C}_2 - \mathbf{C}_1)\,\mathbf{H}\,\mathbf{X}_2
        \;+\;
        f_1\,\mathbf{C}_1\,\boldsymbol{\alpha}_1
        \;+\;
        f_2\,\mathbf{C}_2\,\boldsymbol{\alpha}_2
    \Bigr],
\end{equation}
where the $\mathbf{X}_1$ and $\mathbf{X}_2$ are analytical functions of $(\mathbf{C}_1, \mathbf{C}_2, \alpha_1, \alpha_2, f_1, f_2)$.

Finally, the thermomechanical DMN was validated through thermo-elastic-viscoplastic simulations, where mechanical boundary conditions were applied while temperature variations were progressively introduced. The results confirmed that the model accurately captures the coupled mechanical and thermal responses of heterogeneous materials, demonstrating its effectiveness in thermomechanical analysis.

\subsubsection{Thermal DMN} 

\begin{figure}[H]
    \centering
    \includegraphics[width=0.7\textwidth]{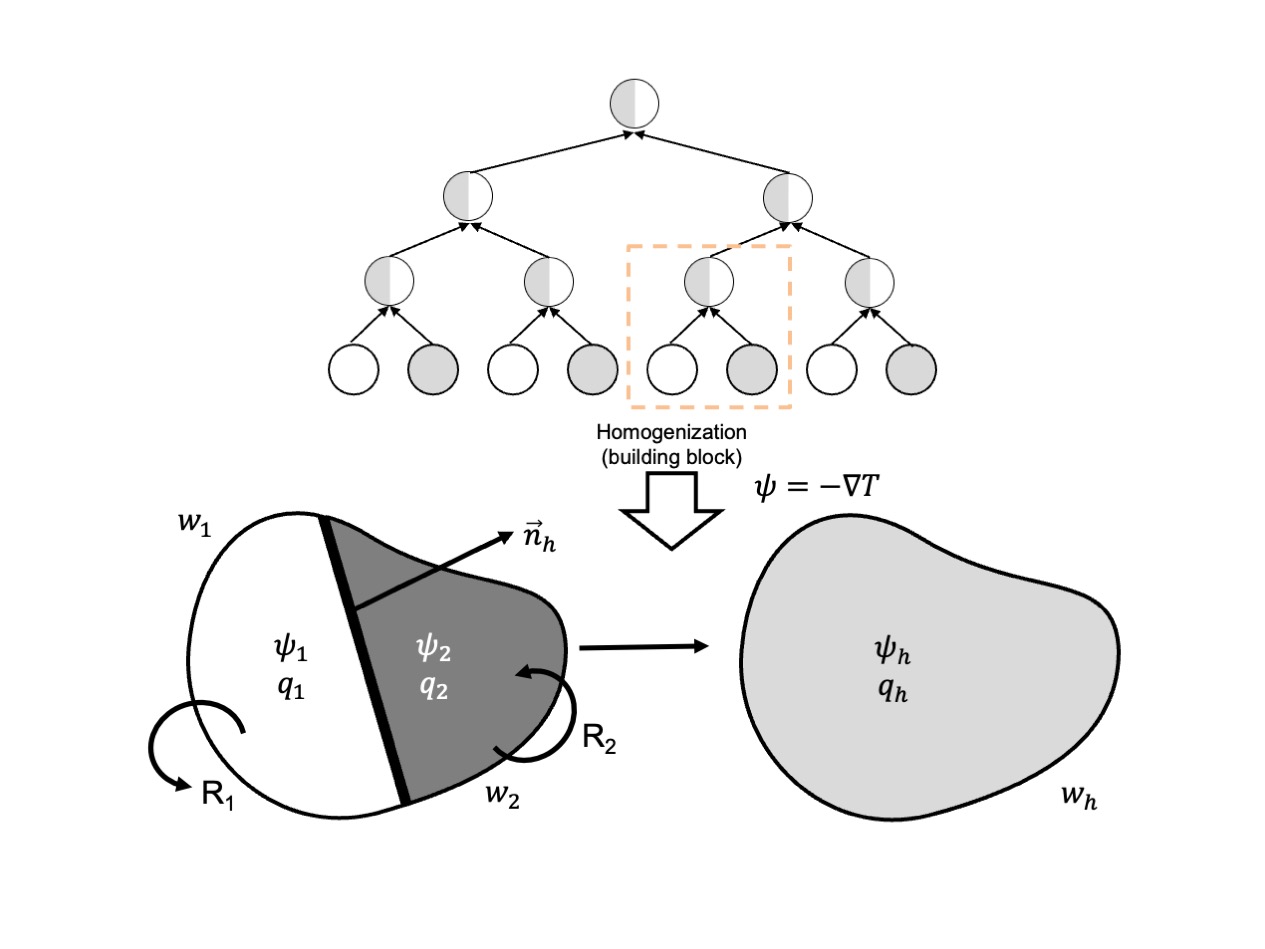}
    \caption{ Schematic illustration of thermal DMN framework.
    }
    \label{fig:thermal conductivity}
\end{figure}

Shin et al. proposed the thermal DMN to model thermal conductivity in woven composite materials~\cite{shin2024deep2}. This framework explicitly incorporates both the heat flux $\textbf{q}$ and the temperature gradient $\psi = -\nabla T$ into the building block formulation, enabling an accurate representation of thermal transport in anisotropic and heterogeneous materials, as shown in Fig.~\ref{fig:thermal conductivity}.

Within each building block, the heat flux vectors for phase 1, phase 2, and the homogenized response are denoted as $\textbf{q}_1$, $\textbf{q}_2$, and $\textbf{q}_h$, respectively, while the corresponding temperature gradient vectors are $\psi_1$, $\psi_2$, and $\psi_h$. The formulation is governed by the following conditions:

\begin{itemize}
    \item Entropy convergence condition:
\begin{equation}\label{eq:entropy} \psi_1 - \psi_2 = \zeta \vec{n}_h / (f^1 f^2) \end{equation} where $\vec{n}_h$ represents the interface direction between the two phases.

    \item Homogenization of field variables:
\begin{equation}\label{eq:homogenization1} \psi_h = f^1 \psi_1 + f^2 \psi_2 \end{equation} \begin{equation}\label{eq:homogenization2} \textbf{q}_h = f^1 \textbf{q}_1 + f^2 \textbf{q}_2 \end{equation}

    \item Heat flux continuity condition:
\begin{equation}\label{eq:continuity}
\vec{n}_h^\top(\textbf{q}_2 - \textbf{q}_1) = 0
\end{equation}

    \item Constitutive relations:
\begin{equation}\label{eq:constitutive1} \textbf{q}_1 = \textbf{R}^\top_1 \textbf{k}_1 \textbf{R}_1 \psi_1 \end{equation} \begin{equation}\label{eq:constitutive2} \textbf{q}_2 = \textbf{R}^\top_2 \textbf{k}_2 \textbf{R}_2 \psi_2 \end{equation} \begin{equation}\label{eq:constitutive3} \textbf{q}_h = \textbf{k}_h \psi_h \end{equation}
where $\textbf{R}_1$ and $\textbf{R}_2$ are the trainable rotation matrices associated with the child nodes, allowing the model to capture the anisotropic characteristics of the thermal conductivity tensor adaptively.

\end{itemize}

Each building block consists of 19 variables, including $\textbf{q}_1$, $\textbf{q}_2$, $\textbf{q}_h$, $\psi_1$, $\psi_2$, $\psi_h$, and $\zeta$. These variables are constrained by 19 equations, as given in Eqs.~\eqref{eq:entropy}--\eqref{eq:constitutive3}, which enables the derivation of an analytical homogenization function for thermal conductivity.

Furthermore, the nodal rotation mechanism in thermal DMN enhances its ability to represent directional heat conduction, improving the accuracy of thermal behavior modeling in complex woven composites.

\subsubsection{DMN with damage effect} 
\begin{figure}[H]
    \centering
    \includegraphics[width=0.8\textwidth]{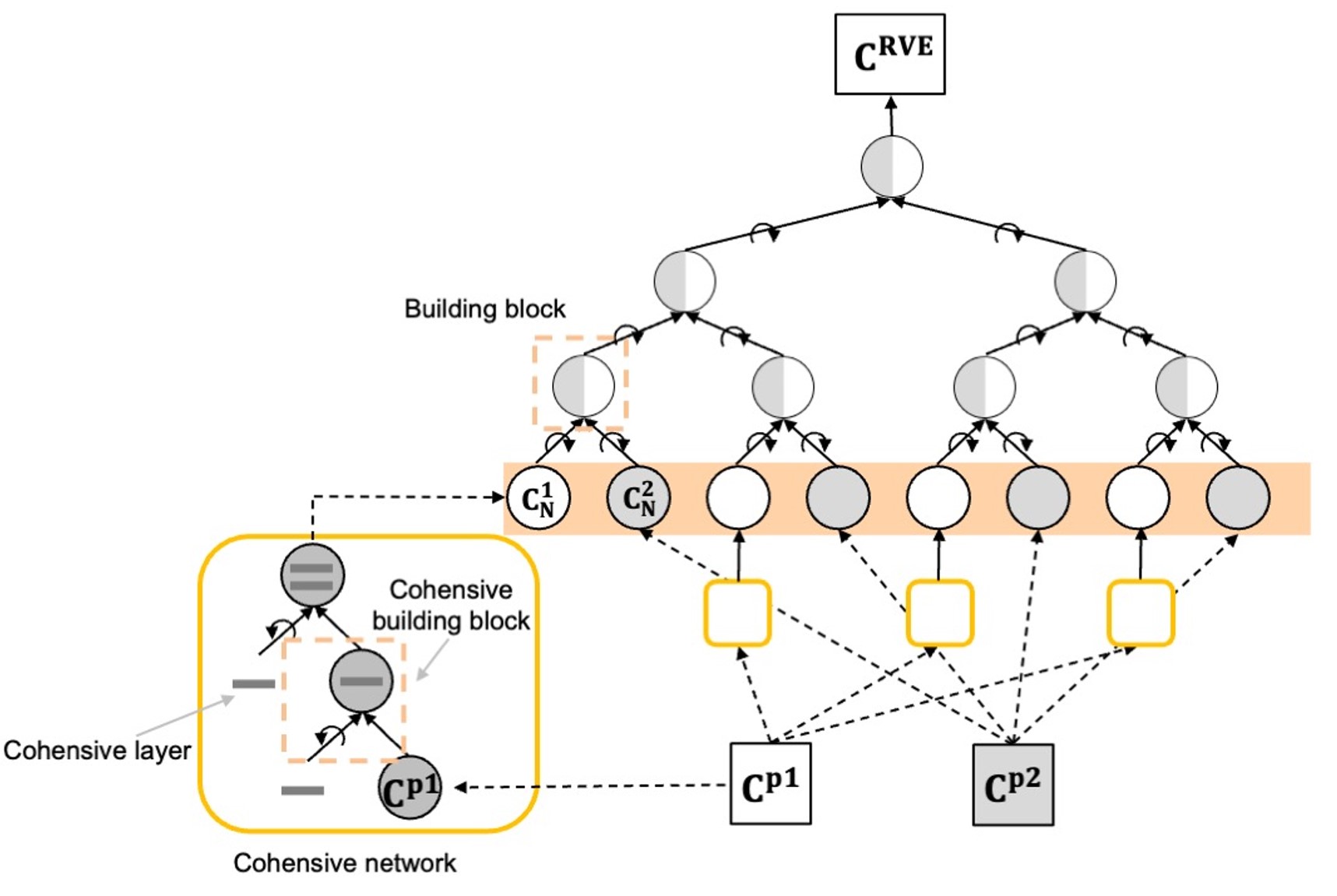}
    \caption{Schematic illustration of the cohesive network integration within DMN.
    }
    \label{fig:cohesive_DMN}
\end{figure}

To model the progressive degradation of materials, the DMN framework can be extended to incorporate interfacial failure and debonding. This is achieved by enriching specific bottom nodes of the DMN and associating them with dedicated cohesive networks. Each cohesive network consists of multiple cohesive layers designed to capture interfacial behavior, as illustrated in Fig.~\ref{fig:cohesive_DMN} \cite{liu2020deep, liu2021cell}.

For each enriched node $q$, the cohesive layers indexed by $p$ are sequentially embedded within the base material through cohesive building blocks. To maintain material consistency, the base material in the cohesive network must represent the same microscale constituent as the one originally present in the bottom layer of the DMN. Each enriched node $q$ is associated with a distinct cohesive network, parameterized by a set of trainable variables: $\left\{ \tilde{z}^p_q, \tilde{\alpha}^p_q, \tilde{\beta}^p_q, \tilde{\gamma}^p_q \right\}$. Additionally, a reciprocal length parameter $\tilde{\nu}$ is introduced to account for the size effects of the interface, defined as:

\begin{equation}
    \tilde{\nu}^p_q = \frac{\text{max}(\tilde{z}^p_q,0)}{L}
\end{equation}
where $L$ is the characteristic length of the RVE.

The traction-separation law governs the response of the cohesive layers and is expressed as:
\begin{equation} 
    \Delta d = G \Delta t + \delta d, 
\end{equation}
where:
\begin{enumerate}
    \item $\Delta d$ represents the incremental separation displacement,

    \item $\Delta t$ represents the incremental separation force,

    \item $G$ is the cohesive layer stiffness matrix,

    \item $\delta d$ is the residual displacement vector.
    
\end{enumerate}

Thus, the homogenized compliance matrix of the cohesive building block can be analytically derived as:

\begin{equation}
    D = D(D^0, G, \tilde{z}, \tilde{\alpha}, \tilde{\beta}, \tilde{\gamma}),
\end{equation}
along with the residual strain formulation:
\begin{equation}
    \delta \varepsilon = \delta \varepsilon (\delta \varepsilon^0, \delta d, \tilde{z}, \tilde{\alpha}, \tilde{\beta}, \tilde{\gamma}).
\end{equation}

Here, \( D^0 \) and \( D \) denote the compliance matrices before and after homogenization, respectively. The parameters \( \tilde{z}, \tilde{\alpha}, \tilde{\beta}, \tilde{\gamma} \) are trainable, while \( \delta \varepsilon^0 \) and \( \delta \varepsilon \) represent the residual strain variables before and after homogenization.

The training process for the DMN with cohesive layers follows a two-stage approach. In the first stage, the DMN is trained independently without incorporating the cohesive network. In the second stage, the cohesive network is integrated while retaining the previously trained DMN parameters. The fully trained model, initially developed using elastic data, can then be extended to capture nonlinear material behavior.

\subsubsection{Flexible DMN } 

The Flexible DMN (FDMN) extends the standard DMN framework to model suspensions of rigid fibers in a non-Newtonian fluid~\cite{sterr2024deep}. A closed-form solution for two-layer linear homogenization is derived, establishing a theoretical foundation for this approach. By incorporating the rheological behavior of the surrounding medium, FDMN effectively predicts the effective stress response of shear-thinning fiber suspensions embedded in a Cross-type matrix material across a wide range of shear rates and under various loading conditions. This formulation enables accurate modeling of the complex interplay between fiber rigidity and non-Newtonian fluid behavior in heterogeneous materials.

\subsection{Generalizing DMNs to diverse microstructures} 

The original DMN framework has a fundamental limitation: a single DMN is typically trained for a specific microstructure. If the microstructure changes, the model must be retrained, which is impractical for industrial applications where microstructural variations are common. To address this challenge, several studies have explored methods to extend DMNs to handle diverse microstructures without requiring complete retraining. These advancements focus on integrating microstructural descriptors, adaptive training strategies, and transferable representations to enhance model generalization. Table \ref{tab:dmn_microstructures} summarizes key developments in this area, highlighting different approaches to improving DMN adaptability across varying microstructures.

\begin{table}[h!]
\centering
\caption{DMN-based models for handling diverse microstructures.}
\label{tab:dmn_microstructures}
\begin{tabular}{@{}llll@{}} % 使用 @{} 去掉內部額外的間距
\toprule
\textbf{Study} & \textbf{Method} & \textbf{Microstructure} & \textbf{Ref} \\
\midrule
Transfer Learning DMN  & Linear Interpolation & Circular Inclusions & \cite{liu2019transfer} \\
MgDMN & Linear Interpolation & 3D SFRP\footnotemark[1] & \cite{huang2022microstructure} \\
MIpDMN & Linear Interpolation & Circular Inclusions & \cite{li2024micromechanics} \\
GNN-DMN & GNN & Short Fibers & \cite{jean2024graph} \\
FM-IMN & FM & Short Fibers & \cite{wei2025foundation} \\
CNN-IMN & CNN & Unidirectional Fibers & \cite{wu2026convolutional} \\
\bottomrule
\end{tabular}
\footnotetext[1]{SFRP: Short Fiber Reinforced Polymer.}
\end{table}

\newpage

\subsubsection{Simple interpolation} 

A straightforward approach to adapting a pre-trained DMN for new microstructures involves parameter interpolation~\cite{liu2019transfer}. Since the trainable weights in the DMN reflect the volume fraction of the material phases, the model can be adjusted for a new microstructure by rescaling its parameters. Consider a pre-trained DMN with a phase 2 volume fraction $vf^{\text{trained}}_2$, while an unseen RVE has a different but known volume fraction $vf^{\text{new}}_2$. The parameters at the bottom node indexed by $j$ are given by:
\begin{equation}
    z^j_{\text{unseen}} =
\begin{cases}
\frac{1-vf^{\text{new}}_2}{1-vf^{\text{trained}}_2}\text{ReLu}(z^j_{\text{trained}}), & \text{if } j \text{ is odd}, \\
\frac{vf^{\text{new}}_2}{vf^{\text{trained}}_2}\text{ReLu}(z^j_{\text{trained}}), & \text{if } j \text{ is even}.
\end{cases}
\end{equation}

The trainable rotation angles for each building block remain unchanged:

\begin{equation}
    \begin{cases}
\alpha^k_{i,{\text{unseen}}} = \alpha^k_{i,{\text{trained}}}, \\
\beta^k_{i,{\text{unseen}}} = \beta^k_{i,{\text{trained}}}, \\
\gamma^k_{i,{\text{unseen}}} = \gamma^k_{i,{\text{trained}}}.
\end{cases}
\end{equation}

While simple parameter rescaling enables the adaptation of pre-trained DMNs to different phase volume fractions without additional training, its accuracy is fundamentally limited. This approach assumes the microstructural interaction geometry remains invariant, an assumption that fails under strong nonlinearities such as plasticity, creep, or fatigue. In these regimes, localized nonlinear effects induce non-uniform microstructural changes that a simple a priori rescaling cannot capture.

To address these limitations, Dey et al. developed a framework that represents microstructural variability through a discrete set of base networks~\cite{dey2024effectiveness}. By training 15 individual DMNs at representative points across the planar fiber-orientation triangle, the model captures a broader range of states. During macroscopic simulations, each Gauss point evaluates the three DMNs corresponding to its specific orientation and determines the final stress response via a convex combination of their predictions. This a posteriori stress-level interpolation significantly enhances accuracy and robustness for highly anisotropic and nonlinear loading conditions.

\subsubsection{Transfer learning} 
To improve upon interpolation-based adaptation, transfer learning has been explored as a means of extending DMNs to new microstructures while reducing the need for full retraining~\cite{liu2019transfer}.

In this approach, a dataset is first generated with multiple predefined volume fractions (e.g., $vf_2 = 0.1, 0.2,..., 0.6$), and corresponding DMN models are trained. If an unseen RVE has a volume fraction $vf^{\text{new}}_2$ that falls within the range [$vf^{\text{low}}_2$, $vf^{\text{high}}_2$], its DMN parameters are interpolated from the nearest pre-trained models as

\begin{equation}
    z^j_{\text{unseen}} = \rho \text{ReLu}(z^j_{\text{low}})
 + (1-\rho)\text{ReLu}(z^j_{\text{high}})
\end{equation}
where the interpolation coefficient $\rho$ is defined as: 
\begin{equation}
    \rho =\frac{vf^{\text{high}}_2 - vf^{\text{new}}_2}{vf^{\text{high}}_2 - vf^{\text{low}}_2}.
\end{equation}

Similarly, the trainable rotation angles at depth $i$ and index $k$ are interpolated as:
\begin{equation}
        \begin{cases}
\alpha^k_{i,{\text{unseen}}} = \rho\alpha^k_{i,{\text{low}}}+(1-\rho)\alpha^k_{i,{\text{high}}}, \\
\beta^k_{i,{\text{unseen}}} = \beta^k_{i,{\text{low}}}+(1-\rho)\beta^k_{i,{\text{high}}},, \\
\gamma^k_{i,{\text{unseen}}} = \gamma^k_{i,{\text{low}}}+(1-\rho)\gamma^k_{i,{\text{high}}},.
\end{cases}
\end{equation}

This transfer-learning-based interpolation strategy enables DMNs to generalize to new microstructures more effectively than simple interpolation, leveraging a broader range of pre-trained models. By integrating knowledge from multiple microstructures, this approach mitigates errors arising from simple parameter scaling and provides more accurate predictions across a diverse range of material configurations.

\subsubsection{Microstructure-guided DMN } 
\begin{figure}[H]
    \centering
    \includegraphics[width=1.0\textwidth]{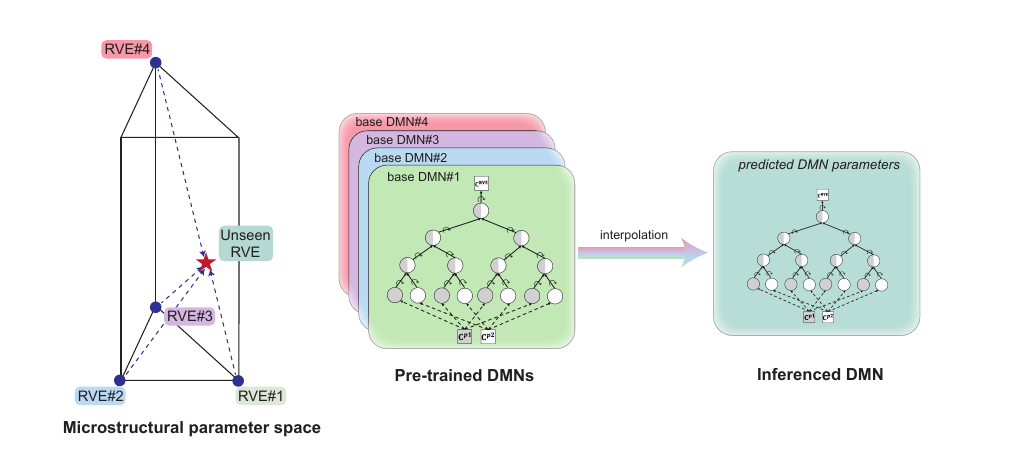}
    \caption{
    Schematic illustration of the microstructural parameter space. The DMN parameters for an unseen RVE are obtained by interpolating four base pre-trained DMNs in this space.
    }
    \label{fig:MgDMN}
\end{figure}

For complex RVE geometries, such as SFRP, the microstructure-guided DMN (MgDMN) extends the standard DMN framework by introducing a parameterization of the microstructure space~\cite{huang2022microstructure}. This enhancement enables MgDMN to capture critical microstructural features and generalize across a broad range of microstructures, as shown in Fig.~\ref{fig:MgDMN}. Unlike traditional DMNs that must be retrained for every new microstructure, MgDMN employs an interpolation scheme to infer DMN parameters using only a few pre-trained DMNs.

In MgDMN, the DMN homogenization process considers both the volume fraction (VF) and the orientation state. For a building block $\mathcal{B}^k_i$ at layer $i$ and index $k$, the VF (denoted by $\text{VF}^k_i$) is defined as:
\begin{equation}
    \text{VF}^k_i = f^{1}\text{VF}^{2k-1}_{i+1} + f^{2}\text{VF}^{2k}_{i+1}
\end{equation}
where $\text{VF}^k_i$ is the actual volume fraction of phase 1, and $f^1$ and $f^2$ are the weight proportions of the child building blocks (computed via Eq.~\eqref{eq:vf_cal}).

The orientation state is aggregated based on the dyadic product of a directional vector $\vec{p}$. The second-order orientation tensor $A=[a_{ij}]$ is given by~\cite{advani1987use}:

\begin{equation}
    a_{ij} = \int p_i \otimes p_j \Phi(\vec{p}) d\vec{p}.
\end{equation}
where $\Phi(\vec{p})$ is the probability density function describing the orientation distribution.

Within this framework, the microstructure is characterized by the VF and orientation tensors. After considering rotational symmetries, the parameter space can be reduced further. In general, for an $N_p$-phase microstructure, $6N_p-8$ distinct microstructures are required to span the microstructural parameter space fully. In the special case of a two-phase SFRP, four distinct microstructures suffice to span the prism-shaped parameter space $\{\nu, a_{11}, a_{22}, a_{33} \}$. 

These four microstructures correspond to:
\begin{enumerate}
    \item Randomly oriented fibers at an $L\%$ fiber volume fraction,
    \item Planar random fibers at an $L\%$ fiber volume fraction,
    \item Aligned fibers at an $L\%$ fiber volume fraction,
    \item Aligned fibers at an $H\%$ fiber volume fraction.
\end{enumerate}
Here, $L$ and $H$ are set to 10\% and 15\%, respectively. Separate DMNs, referred to as the base DMNs, are trained on each of these four RVEs. For an unseen microstructure, interpolation is used to determine DMN parameters. Specifically, the unseen microstructure is projected into the parameter space to obtain $\left\{ \tilde{\nu}, \tilde{a}_{11}, \tilde{a}_{22}, \tilde{a}_{33} \right\}$. The volume percentages $\{w_1, w_2, w_3, w_4\}$ associated with each base DMN are then computed by solving:

 \begin{equation}
     \begin{bmatrix}
a^{(1)}_{11} & a^{(2)}_{11} & a^{(3)}_{11} & a^{(4)}_{11} \\
a^{(1)}_{22} & a^{(2)}_{22} & a^{(3)}_{22} & a^{(4)}_{22} \\
a^{(1)}_{33} & a^{(2)}_{33} & a^{(3)}_{33} & a^{(4)}_{33} \\
\nu^{(1)} & \nu^{(2)} & \nu^{(3)} & \nu^{(4)} \\
\end{bmatrix}
\begin{bmatrix}
w^{(1)} \\
w^{(2)} \\
w^{(3)} \\
w^{(4)} \\
\end{bmatrix}=
\begin{bmatrix}
\tilde{a}_{11} \\
\tilde{a}_{22} \\
\tilde{a}_{33} \\
\tilde{\nu} \\
\end{bmatrix}.
 \end{equation}
Here, the superscript $^{(i)}$ represents the index of the base DMN microstructure parameters. Once these weights are determined, the DMN parameters $\tilde{\mathcal{P}}$ for the unseen microstructure are obtained as a weighted linear combination of the base DMNs:

\begin{equation}
    \tilde{\mathcal{P}} = \sum_{i=1}^{4} w^{(i)} \mathcal{P}^{(i)}
\end{equation}
where $\mathcal{P}$ represents DMN parameter (e.g., $z^{k=1,...,2^{N-1}}$, $\alpha ^{k=1,...,2^{i-1}}_{i=1,...,N}$, $\beta ^{k=1,...,2^{i-1}}_{i=1,...,N}$, or $\gamma ^{k=1,...,2^{i-1}}_{i=1,...,N}$.

\subsubsection{Micromechanics-informed parametric DMN} 

\begin{figure}[H]
    \centering
    \includegraphics[width=0.7\textwidth]{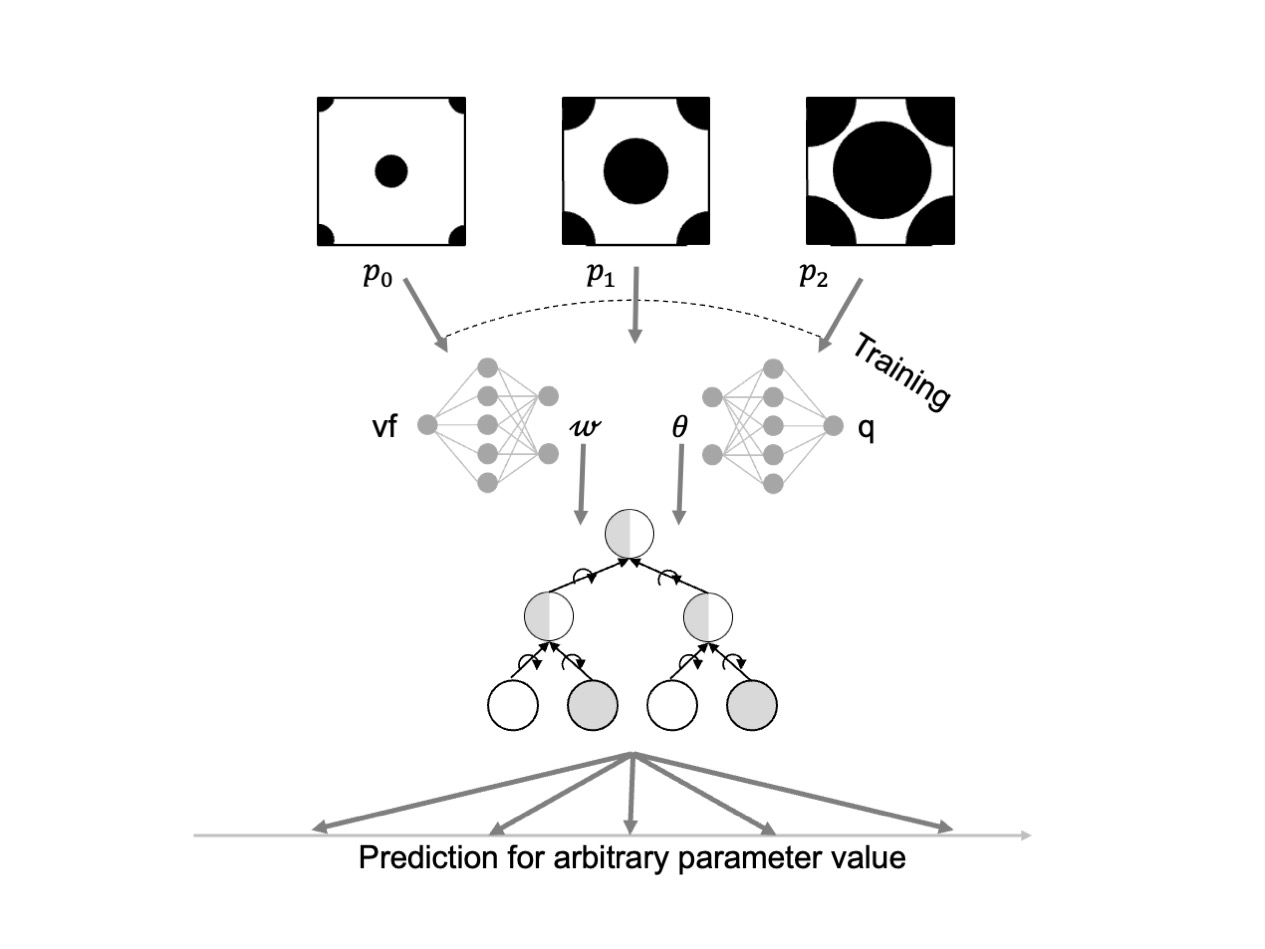}
    \caption{ Schematic illustration of the MIpDMN framework.
    }
    \label{fig:MIpDMN}
\end{figure}

The micromechanics-informed parametric DMN (MIpDMN) incorporates micromechanical parameters to accommodate microstructural variations across different representative volume elements (RVEs) \cite{li2024micromechanics}. In MIpDMN, each microstructure is projected onto a parametric space $\vec{p}$, as shown in Fig.~\ref{fig:MIpDMN}: 
\begin{equation} 
    \vec{p} = \{ v_f, \vec{q} \} \in \mathbb{R} \times \mathbb{R}^q 
\end{equation} 
where $v_f$ is the volume fraction, and $\vec{q}$ represents morphological parameters orthogonal to $v_f$. 

Since DMN parameters inherently encode microstructural characteristics, they can be expressed as functions of $\vec{p}$. Consequently, the overall homogenization process becomes:
\begin{equation}
    (\vec{p},\textbf{C}^{p1}, \textbf{C}^{p2})\mapsto \bar{\textbf{C}}
\end{equation}
where $\textbf{C}^{p1}$ and $\textbf{C}^{p2}$ denote the phase stiffness matrices, and $\bar{\textbf{C}}$ is the homogenized stiffness matrix.

The trainable DMN parameters can be classified into volume fraction-dependent parameters and rotation angles. In MIpDMN, the weight $w$ is redefined as a linear function of volume fraction with the activation function $\sigma$: 
\begin{equation}\label{eq:MIpDMN1} 
    w(\vec{p}) = w(v_f) = \sigma(W_1 \cdot v_f + W_0) 
\end{equation}

Similarly, the DMN rotation angles $\theta$ are modeled as a linear transformation of $\vec{q}$:
\begin{equation}\label{eq:MIpDMN2}
    \theta(\vec{p})  = \theta(\vec{q}) = \text{ReLu}(\Theta_1 \cdot \vec{q} + \Theta_0)
\end{equation}
where $\theta$ represents the set of rotation angles $\{ \alpha, \beta, \gamma \}$ in the DMN.

In summary, the trainable parameters in MIpDMN related to the volume fraction consist of $2^{N+1}$ parameters, including $w^j$, $W_1$, and $W_0$. Meanwhile, the trainable parameters orthogonal to the volume fraction amount to $4 \times (q+1)\times(2^{N}-1) $. For an unseen RVE, its microstructural representation is first mapped onto the parametric space $\vec{p}$, after which Eq.\eqref{eq:MIpDMN1} and Eq.\eqref{eq:MIpDMN2} are used to determine the corresponding MIpDMN parameters.

\subsubsection{GNN-DMN framework} 
Recent advancements in DMNs have been achieved by integrating graph neural networks (GNNs) into the DMN framework, leading to the development of the GNN-DMN approach \cite{jean2024graph}. In GNN-DMN, the microstructure is discretized into a mesh, which is then used to construct a graph $\mathbb{G}$. Each mesh element is treated as a node, with mesh connectivity defining the graph edges. The node attributes include:

\begin{enumerate}
    \item Area of mesh element
    \item xy-coordinates of the element centroid
    \item Phase of the mesh element
    \item Boundary status of the element (whether it lies on the boundary) 
\end{enumerate}

These attributes are processed by a GNN to extract a latent vector $X_{\text{feat}}$, which is subsequently transformed into the DMN parameters $\vec{p}$ via a fully connected network. The homogenization process in GNN-DMN is formulated as:
\begin{equation}
    (\vec{p} (\mathbb{G}),\textbf{C}^{p1}, \textbf{C}^{p2})\mapsto \bar{\textbf{C}}
\end{equation}
where $\vec{p}$ consists of the following DMN parameters:
\begin{equation}
    \vec{p}=\{z^{k=1,...,2^{N-1}}, \alpha ^{k=1,...,2^{i-1}}_{i=1,...,N}, \beta ^{k=1,...,2^{i-1}}_{i=1,...,N}, \gamma ^{k=1,...,2^{i-1}}_{i=1,...,N} \}
\end{equation}

\begin{figure}[H]
    \centering
    \includegraphics[width=1.0\textwidth]{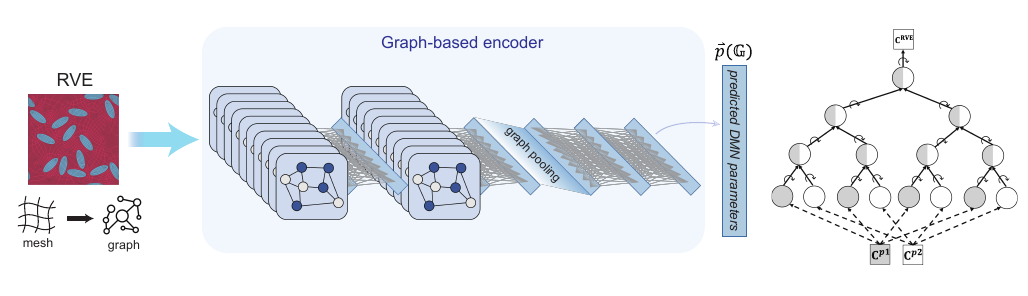}
    \caption{
        Schematic illustration of the GNN-DMN framework.
    }
    \label{fig:GNN-DMN}
\end{figure}

For an unseen RVE, the microstructure is first discretized into a mesh and represented as a graph $\mathbb{G}$. The trained GNN-DMN framework then extracts the corresponding DMN parameters $\vec{p}$, enabling homogenization without requiring additional parameter fitting, as shown in Fig.~\ref{fig:GNN-DMN}.

\subsubsection{Foundation model-IMN framework} 

Foundation models (FMs) have achieved remarkable success across various domains, including natural language processing, computer vision, and, more recently, composite materials modeling \cite{wei2025foundation}. In the context of material modeling, a recent study introduced an FM-based approach that leverages a masked autoencoder (MAE) to learn microstructural representations from a large dataset of 100,000 composite microstructures.

The MAE employs an encoder-decoder architecture for self-supervised learning. During pretraining, a portion of the input microstructure image is masked, and the encoder extracts latent features from the unmasked regions. The decoder then reconstructs the missing parts, compelling the encoder to capture essential structural patterns and dependencies. Through this process, the encoder learns a low-dimensional representation that effectively encodes microstructural information while remaining robust to missing data.

\begin{figure}[H]
    \centering
    \includegraphics[width=1.0\textwidth]{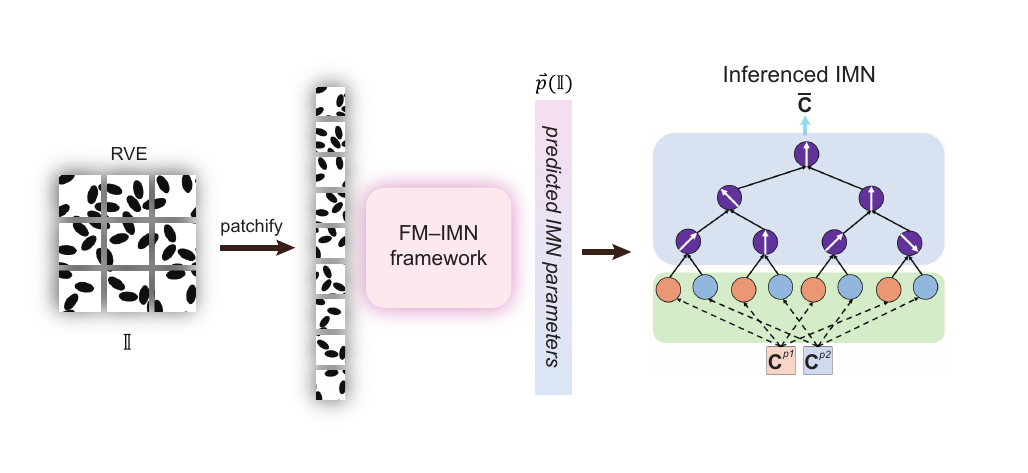}
    \caption{
        Schematic illustration of the FM-IMN framework.
    }
    \label{fig:FM-IMN}
\end{figure}

Once pre-trained, the MAE encoder is extracted and repurposed for downstream tasks. In the FM-IMN framework, a linear projection layer is appended to the pre-trained encoder, directly mapping microstructural images to IMN parameters, as shown in Fig.~\ref{fig:FM-IMN}. This transfer learning approach enables the foundation model to be fine-tuned for homogenization tasks. The homogenization process is formulated as:
\begin{equation}
    (\vec{p}(\mathbb{I}),\textbf{C}^{p1}, \textbf{C}^{p2})\mapsto \bar{\textbf{C}}
\end{equation}
where $\mathbb{I}$ is the grayscale image of the microstructure, and $\vec{p}(\mathbb{I})$ represents the predicted IMN parameters.

For an unseen microstructure, its grayscale image is processed by the fine-tuned FM-IMN framework. Since this framework eliminates the need for explicit parameter fitting for each new microstructure, it enables efficient nonlinear extrapolation across diverse material systems.

\subsubsection{Comparative Summary of Generalization Strategies} 

The evolution of DMN generalization strategies reflects a clear trade-off between {computational simplicity} and \textbf{representational capacity}. As summarized in Table \ref{tab:dmn_microstructures}, these methodologies can be categorized into three distinct levels of complexity based on their treatment of the microstructural design space:

\begin{itemize}
    \item {Interpolation-based methods} (Simple Interpolation, Transfer Learning, and MgDMN) leverage the inherent multilinearity of the DMN architecture. These are highly efficient and physically intuitive, as they utilize the volume fraction or orientation tensors to rescale the weights and angles of base networks. However, their utility is largely confined to microstructural families with well-defined, low-dimensional parameter spaces, such as circular inclusions or short-fiber reinforced polymers.
    
    \item {Parametric models} (MIpDMN) enhance flexibility by treating rotation angles and weights as functional mappings of morphological parameters. While this approach extends the model's reach to microstructures orthogonal to volume fraction, it remains dependent on the manual identification of relevant descriptors, which can be challenging for disordered or irregular RVEs.
    
    \item {Representation learning frameworks} (GNN-DMN and FM-IMN) represent the current state-of-the-art in microstructural generalizability. By integrating Graph Neural Networks or Masked Autoencoders, these models bypass the need for manual feature engineering. They extract high-dimensional latent vectors directly from raw mesh or image data to predict DMN parameters. Although they require substantial offline datasets and computational resources for pre-training, they offer a robust pathway toward "foundation models" for mechanics that can handle arbitrary microstructural topologies without retraining.
\end{itemize}

In summary, selecting a DMN generalization strategy involves balancing the available training data with the required geometric diversity. For standardized industrial composites, MgDMN provides a high-efficiency solution with minimal data overhead. Conversely, for advanced materials discovery involving complex, non-periodic, or evolving geometries, the FM-IMN and GNN-DMN frameworks provide the necessary representational power to ensure predictive accuracy across diverse material systems.

\section{Applications}\label{sec4}

DMN has emerged as a powerful surrogate model for two-scale analyses, offering substantial computational efficiency while maintaining accuracy. Initially applied to SFRP, their applicability has since expanded across diverse material systems and industrial applications.

A key application of DMNs is in SFRP component analysis~\cite{gajek2021fe, gajek2021efficient}. By replacing the conventional RVE with a DMN at each Gaussian integration point, the FE–DMN framework enables efficient multiscale simulations within ABAQUS via an implicit user-material (UMAT) subroutine. In a study of a quadcopter drone with over 9 million DOFs, fiber orientation data from injection-molding simulations were incorporated to perform high-fidelity structural analysis. Despite the model's complexity, the simulation completed in 267 minutes using 252 GB of DRAM, demonstrating the feasibility of DMNs for large-scale industrial applications.

DMNs have also been extended to thermomechanically coupled composite materials~\cite{gajek2022fe, gajek2023deep}. For instance, a non-symmetric notched plate under cyclic loading was analyzed using a two-scale simulation to capture the fully coupled thermomechanical response. Compared to conventional approaches, the DMN-based method achieved a computational speedup of five to six orders of magnitude while preserving accuracy. This capability is particularly relevant for applications involving temperature-dependent material behavior.

Another notable application involves impact and contact simulations of composite structures under dynamic loading. One study integrated fiber orientation and volume fraction data from Moldex3D injection molding simulations into a DMN surrogate model, which was then coupled with LS-DYNA for large-scale impact analysis~\cite{wei2023ls}. This approach enables efficient and accurate simulations of dynamic impact events, offering a computationally viable alternative to conventional multiscale modeling methods. The ability of DMNs to handle large-scale impact scenarios makes them particularly valuable for designing lightweight, crashworthy materials in automotive applications.

Furthermore, DMNs have been employed for inverse identification of material parameters in short fiber-reinforced thermoplastics, a process that remains computationally demanding even with FFT-based methods. By replacing full-field simulations with DMN surrogates, this methodology has significantly reduced computational costs while maintaining high accuracy~\cite{dey2023rapid}.

More recently, DMNs have been increasingly integrated into industrial manufacturing workflows through virtual product development (VPD). For example, Meyer et al. employed DMNs in combination with molding process simulations to predict the stochastic effective mechanical performance of sheet molding compound (SMC) composites, enabling probabilistic assessment of process-induced variability~\cite{meyer2023probabilistic}.
Furthermore, Wu et al. reformulated the IMN by decoupling the phase volume fraction from the microstructural parameters, thereby introducing randomness in the phase volume fraction for Stochastic Volume Elements (SVEs) and enabling stochastic nonlinear response prediction~\cite{wu2025stochastic}.
In a complementary direction, Robertson et al. developed the Variational Deep Material Network (VDMN), in which variational distributions are embedded into the mechanistic building blocks of the network to explicitly capture microstructure-induced aleatoric uncertainties. This framework enables probabilistic forward and inverse modeling and establishes a foundation for uncertainty-robust materials digital twins~\cite{robertson2025microstructure}.

In summary, DMNs have proven effective in a wide range of multiscale and multiphysics applications, including structural and thermomechanical analysis, impact modeling, and inverse parameter identification. Their flexibility and computational efficiency make them a promising tool for advanced material modeling and industrial applications, with ongoing research continuously expanding their scope.

\section{Conclusions}
This study provides a comprehensive synthesis of the theoretical
foundations of DMNs, tracing their evolution from
fundamental building blocks to advanced frameworks for
multiphysics modeling. By reviewing diverse strategies for
microstructural adaptation and their integration into FE–
DMN frameworks, we have demonstrated the versatility of
DMNs in addressing large-scale structural structural simulations
as well as complex inverse material identification
problems.

The distinctive strength of the DMN framework resides in
its hierarchical architecture, enabling the explicit derivation
of two-phase linear homogenization operators. This structure
enables a highly efficient "offline" training phase using
linear datasets while preserving robust "online" extrapolation
capability into nonlinear regimes. Consequently,
DMNs effectively bridge the gap between rigorous physics-
based homogenization and the high-speed execution of
data-driven surrogates, offering a unique blend of interpretability,
computational efficiency, and physical consistency.

However, the transition from research to widespread
industrial adoption necessitates addressing several critical
bottlenecks identified in this review:
\begin{enumerate}
    \item Multiphysics Integration: While mechanical homogenization within DMNs has reached a relatively mature stage, extending DMNs to fully coupled nonlinear multiphysics phenomena, such as piezoelectricity, electroconductivity, and saturated porous media, remains in its early stages. Leveraging the inherent extensibility of the DMN architecture to address these domains constitutes a key priority for future research.

    \item Scale Separation Limits: Most current formulations are restricted to first-order homogenization and rely on the assumption of strict scale separation. This assumption
limits predictive fidelity in scenarios involving pronounced
size effects or strain localization.

    \item Microstructural Generalizability: A primary limitation
of current DMNs is their topological rigidity, whereby
substantial retraining is typically required for changes
in microstructural geometry. Although a growing body
of literature has begun to address this issue, most existing
research remains confined to narrow and highly specific
classes of microstructures.

\end{enumerate}

In conclusion, while DMNs have already demonstrated strong potential for high-fidelity multiscale modeling at a substantially reduced computational cost, their long-term impact will depend critically on continued advances in multiphysics coupling and geometric flexibility. Progress along these directions is essential for establishing DMNs as a cornerstone of next-generation digital twins and accelerated materials design.

\newpage

\section*{Acknowledgements}

This work is supported by the National Science
and Technology Council, Taiwan, under Grant 111-2221-E-002-054-MY3, 112-2221-E-007-028, and 114-2221-E-002-010-MY3. We are grateful for the computational resources and support from the NTUCE-NCREE Joint Artificial Intelligence Research Center and the National Center of High-performance Computing (NCHC). 

\section*{Declarations}
The authors declare no conflict of interest.

\section*{Author Contributions}
TJW led the writing of the manuscript and developed the core content. WNW contributed to writing and assisted with figure editing. CSC supervised the research and revised the manuscript to its final form.

\section*{Data Availability} 
No datasets were generated or analysed during the
current study.

\bibliographystyle{unsrt}  
%\bibliography{references}  %%% Remove comment to use the external .bib file (using bibtex).
%%% and comment out the ``thebibliography'' section.
\bibliography{references}

%%% Comment out this section when you \bibliography{references} is enabled.
% \begin{thebibliography}{1}

% \bibitem{kour2014real}
% George Kour and Raid Saabne.
% \newblock Real-time segmentation of on-line handwritten arabic script.
% \newblock In {\em Frontiers in Handwriting Recognition (ICFHR), 2014 14th
%   International Conference on}, pages 417--422. IEEE, 2014.

% \bibitem{kour2014fast}
% George Kour and Raid Saabne.
% \newblock Fast classification of handwritten on-line arabic characters.
% \newblock In {\em Soft Computing and Pattern Recognition (SoCPaR), 2014 6th
%   International Conference of}, pages 312--318. IEEE, 2014.

% \bibitem{hadash2018estimate}
% Guy Hadash, Einat Kermany, Boaz Carmeli, Ofer Lavi, George Kour, and Alon
%   Jacovi.
% \newblock Estimate and replace: A novel approach to integrating deep neural
%   networks with existing applications.
% \newblock {\em arXiv preprint arXiv:1804.09028}, 2018.

% \end{thebibliography}

\end{document}